\documentclass[floatfix, preprint, onecolumn, prd,showpacs, showkeys, preprintnumbers, nofootinbib,hyper, superscriptaddress]{revtex4-1}
\usepackage{times}
\usepackage{amssymb,amsbsy,amsmath,amsfonts}
\usepackage{graphicx}
\usepackage{float}
\usepackage{color}
\usepackage{morefloats}
\usepackage{rotating}
\usepackage{srcltx}
\usepackage{slashed}
\usepackage{multirow}
\usepackage{verbatim}
\usepackage{tabularx}
\renewcommand{\arraystretch}{1.2}

\begin{document}

\title{Dynamically generated $J^P=1/2^-(3/2^-)$ singly charmed and bottom heavy baryons}

\author{Jun-Xu Lu}
\affiliation{School of Physics and Nuclear Energy Engineering and International Research Center for Nuclei and Particles
in the Cosmos, Beihang University, Beijing 100191, China}

\author{Yu Zhou}
\affiliation{School of Physics and Nuclear Energy Engineering and International Research Center for Nuclei and Particles
in the Cosmos, Beihang University, Beijing 100191, China}

\author{Hua-Xing Chen}
\affiliation{School of Physics and Nuclear Energy Engineering and International Research Center for Nuclei and Particles
in the Cosmos, Beihang University, Beijing 100191, China}

\author{Ju-Jun Xie}
\affiliation{Institute of Modern Physics, Chinese Academy of Sciences, Lanzhou 730000, China}

\author{Li-Sheng Geng}
\email[E-mail me at: ]{lisheng.geng@buaa.edu.cn}
\affiliation{School of Physics and Nuclear Energy Engineering and International Research Center for Nuclei and Particles
in the Cosmos, Beihang University, Beijing 100191, China}

\begin{abstract}
Approximate heavy-quark spin and flavor symmetry and chiral symmetry
play an important role in our understanding of the nonperturbative regime of strong interactions. In this work, utilizing the unitarized chiral perturbation theory, we
explore the consequences of these symmetries in the description of the interactions between the ground-state singly charmed (bottom) baryons and the pseudo-Nambu-Goldstone bosons.
In particular, at leading order in the chiral expansion, by fixing the only parameter in the theory to reproduce the $\Lambda_b(5912)$ [$\Lambda_b^*(5920)$] or the $\Lambda_c(2595)$ [$\Lambda_c^*(2625)$], we predict a number of dynamically generated states, which are
contrasted with
those of other approaches and available experimental data. In anticipation of future lattice QCD simulations, we calculate the corresponding scattering lengths and compare them to the existing
predictions from a $\mathcal{O}(p^3)$ chiral perturbation theory study.  In addition, we estimate the effects of the next-to-leading-order potentials by adopting heavy-meson Lagrangians and fixing the relevant low-energy constants using either symmetry or naturalness arguments. It is shown that higher-order potentials play a relatively important role in many channels, indicating that further studies are needed once more experimental or lattice QCD data become available.
\end{abstract}

\pacs{12.39.Fe, 11.10.St,14.20.Mr,  14.20.Lq	}
\keywords{Chiral Lagrangians,  Bound and unstable states; Bethe-Salpeter equations,  Bottom baryons,Charmed baryons}

\date{\today}

\maketitle
\section{Introduction}
In recent years, heavy-flavor hadron physics has yielded many surprising results and attracted a lot of attention due to intensive
worldwide  experimental activities, such as \textit{BABAR}~\cite{Bernard:2013zwa}, Belle~\cite{SANTEL:2013jua,Kato:2014nga}, CLEO~\cite{Seth:2011qd}, BES~\cite{Collaboration):2014uxa}, LHCb~\cite{Simone:2014yla}, and CDF~\cite{Palni:2012zxa}. The discoveries and confirmations
of the many $XYZ$ particles have established the existence of exotic mesons made of four quarks, such as
the $Z_c(3900)$~\cite{Ablikim:2013mio,Liu:2013dau} and the $Z(4430)$~\cite{Aaij:2014jqa,Choi:2007wga}, and aroused great interest in the
theoretical and lattice QCD community to understand their nature, though no consensus has been reached yet (see, e.g.,  Ref.~\cite{Brambilla:2010cs}).

Different from the case of heavy-meson states, no similar exotic states have been firmly established in the heavy-flavor baryon sector, partly due to the fact that their production is more difficult. Up to now, there have only been a few experimental observations of excited charmed and bottom baryons (see Ref.~\cite{Crede:2013kia} for
a recent and comprehensive review).  In the bottom baryon sector, the LHCb Collaboration has reported two  excited  $\Lambda_b$ states, the $\Lambda_b(5912)$ and the $\Lambda_b(5920)$~\cite{Aaij:2012da}, with the latter
being recently confirmed by the CDF Collaboration~\cite{Aaltonen:2013tta}. In the charmed baryon sector,  a number of  excited states have been
confirmed by various experiments, including the $\Lambda_c(2595)$, the $\Xi_c(2790)$, the $\Lambda_c(2625)$, and the $\Xi_c(2815)$~\cite{Agashe:2014kda}. The spin parities of the first two states and the last two states are assumed to be
$1/2^-$ and $3/2^-$, respectively, according to quark model predictions.

The conventional picture is that these states are  the orbital excitations of the corresponding ground states.  There are, however, different
interpretations; namely, they are dynamically generated states from the interactions between the ground-state charmed (bottom) baryons with the pseudo-Nambu-Goldstone bosons  (and other coupled channels)~\cite{Lutz:2003jw,GarciaRecio:2008dp,GarciaRecio:2012db,Liang:2014eba,Liang:2014kra} .
 The idea of dynamically generated states is an old one but  has recently received a lot of attention. It has been quite successful in solving some long-standing
 difficulties encountered in hadron spectroscopy, e.g., the nature of the $\Lambda(1405)$ or the lowest-lying scalar nonet (see, e.g., Ref.~\cite{Hyodo:2011ur} for a recent review).
 In the charmed and bottom baryon sector,~\footnote{It is to be noted that in the charmed mesonic sector,
 many newly observed resonances have been claimed to  of composite nature based on phenomenological Lagrangians or effective field theories~(see, e.g., Refs.~\cite{ Kolomeitsev:2003ac,Guo:2006fu,Gamermann:2006nm,Faessler:2007gv,Ding:2009vj,Dong:2009yp,Bondar:2011ev,HidalgoDuque:2012pq,Li:2012ss,Li:2012mqa,Cleven:2013sq,Altenbuchinger:2013vwa}).} various approaches have been adopted to study final-state interactions and resulting dynamically generated states, including the so-called unitarized chiral perturbation theory (UChPT)~\cite{Lutz:2003jw},
  hidden-gauge symmetry inspired approaches~\cite{ Liang:2014eba,Liang:2014kra,
Wu:2010jy,Wu:2010vk,Wu:2010rv,Xiao:2013yca,Xiao:2013jla}, and heavy-quark symmetry inspired approaches~\cite{GarciaRecio:2008dp,GarciaRecio:2012db,Flynn:2011gf,Romanets:2012hm,Romanets:2012ce,Garcia-Recio:2013gaa,Guo:2013xga} .

 In the present work, we choose the UChPT to study the interactions between the ground-state charmed (bottom) baryons and the pseudo-Nambu-Goldstone bosons  using the
 leading order (LO) chiral Lagrangians. In the charmed baryon case, our study differs from that of Ref.~\cite{Lutz:2003jw} in the following respects.
 First, we adopt different regularization schemes to regularize the loop function in the UChPT. Second, to identify dynamically generated states,
  we search for poles on the complex plane instead of examining speed plots. Furthermore, we extend the UChPT to study the bottom baryons and  study the effects of next-to-leading-order (NLO)
  potentials.

The paper is organized as follows. In Sec. II we briefly recall the  UChPT in the description of the interactions between the pseudo Nambu-Goldstone mesons and the ground-state singly
charmed (bottom) baryons at LO.  Our main results are presented in Sec. III.  In Sec. IV, we perform an exploratory NLO study,  followed by a short summary and outlook in Sec. IV.

\section{Theoretical framework}
In this section, we briefly recall the essential ingredients of the UChPT. There are two building blocks in the UChPT: a kernel provided by chiral Lagrangians up to a certain order and a unitarization procedure. The kernel is standard except in the sector where baryons or heavy hadrons are involved, where nonrelativistic chiral Lagrangians are frequently used. Common unitarization procedures include the Bethe--Salpeter equation method~\cite{  Kaiser:1995eg,Oller:1997ti,Oset:1997it,Krippa:1998us,Nieves:1999bx,Meissner:1999vr,Lutz:2001yb,GarciaRecio:2002td,Hyodo:2002pk,Borasoy:2006sr}, the numerator/denominator (N/D) method~\cite{Oller:1998zr}, and the inverse amplitude method~\cite{  Truong:1988zp,Dobado:1989qm,Dobado:1992ha,Dobado:1996ps, Guerrero:1998ei,GomezNicola:2001as,Pelaez:2004xp}. In the present work, we choose to work with relativistic chiral Lagrangians and in the Bethe--Salpeter equation framework.

The Bethe--Salpeter equation can be written schematically as
\begin{equation}
T=V+VGT,
\end{equation}
where $T$ is the unitarized amplitude, $V$ is the potential, and $G$ is the one-loop two-point scalar function. In the context of the UChPT,
the integral Bethe--Salpeter equation is often simplified and approximated to be an algebraic equation with the use of the on-shell approximation~\cite{Oller:1997ti,Oset:1997it}.
This approximations works very well. See Ref.~\cite{Altenbuchinger:2013gaa} for a recent study of off-shell effects in the UChPT and early references on this subject.

The leading-order interaction between a singly charmed baryon of the ground-state sextet and antitriplet and a pseudoscalar meson of the pion octet is provided
by the chiral Lagrangian~\cite{Lutz:2003jw,Liu:2012uw}
\begin{eqnarray}\label{Eq:LOLag}
        \begin{split}
           \mathcal{L}
           & =\frac{i}{16f_0^{2}}\mathrm{Tr}(\bar{H}_{[\bar{3}]}(x)\gamma^{\mu}[H_{[\bar{3}]}(x),[\phi(x),(\partial_{\mu}\phi(x))]_{-}]_{+})\\
           &  +\frac{i}{16f_0^{2}}\mathrm{Tr}(\bar{H}_{[6]}(x)\gamma^{\mu}[H_{[6]}(x),[\phi(x),(\partial_{\mu}\phi(x))]_{-}]_{+}),
        \end{split}
\end{eqnarray}
where $f_0$ is the pseudoscalar decay constant in the chiral limit, $\phi$ collects the pseudoscalar octet, and $H_{[\overline{3}]}$ and $H_{[6]}$ collect the charmed (bottom) baryons, respectively,
\begin{equation}
   \phi=\sqrt{2}
   \left(
   \begin{array}{ccc}
      \frac{\pi^{0}}{\sqrt{2}}+\frac{\eta}{\sqrt{6}} & \pi^{+} & K^{+} \\
      \pi^{-} & -\frac{\pi^{0}}{\sqrt{2}}+\frac{\eta}{\sqrt{6}} & K^{0} \\
      K^{-} & \overline{K^{0}} & -\frac{2}{\sqrt{6}}\eta
   \end{array}
   \right),
\end{equation}
\begin{equation}
   H_{[\overline{3}]}=
   \left(
   \begin{array}{ccc}
      0 & \Lambda_{c}^{+} & \Xi_{c}^{+} \\
      -\Lambda_{c}^{+} & 0 & \Xi_{c}^{0} \\
      -\Xi_{c}^{+} & -\Xi_{c}^{0} & 0
   \end{array}
   \right),
\end{equation}
\begin{equation}
   H_{[6]}=
   \left(
   \begin{array}{ccc}
      \Sigma_{c}^{++} & \frac{\Sigma_{c}^{+}}{\sqrt{2}} & \frac{\Xi_{c}^{'+}}{\sqrt{2}} \\
      \frac{\Sigma_{c}^{+}}{\sqrt{2}} & \Sigma_{c}^{0} & \frac{\Xi_{c}^{'0}}{\sqrt{2}} \\
      \frac{\Xi_{c}^{'+}}{\sqrt{2}} & \frac{\Xi_{c}^{'0}}{\sqrt{2}} & \Omega_{c}^{0}
   \end{array}
   \right).
\end{equation}
The $H$'s for the corresponding ground-state bottom baryons can be obtained straightforwardly by replacing the charm quark content by its bottom counterpart.

Expanding the Lagrangian of Eq.~(\ref{Eq:LOLag}) up to two pseudoscalar fields, one obtains the interaction kernel needed to
describe the $\phi (p_2) B (p_1) \rightarrow \phi (p_4) B (p_3)$  process, where $p_i$'s are the 4-momenta of the respective particles,
\begin{eqnarray}\label{Eq:LOKernel}
   V=\frac{C_{ij}^{(I,S)}}{4f_0^{2}}\gamma^{\mu}(p_{2}^{\mu}+p_{4}^{\mu})\approx\frac{C_{ij}^{(I,S)}}{4f_0^{2}}(E_2+E_4),
\end{eqnarray}
where $C_{ij}^{(I,S)}$ are the Clebsch--Gordan coefficients given in the Appendix. In deriving the final form of $V$, we have assumed that
the 3-momentum of a baryon is small compared to its mass. This is a valid assumption  since in the present study we are only interested in the energy region close to the threshold of the respective coupled channels.

The loop function $G$ in the Bethe--Salpeter equation has the following simple form in four dimensions:
\begin{equation}
   G=i\int\frac{d^{4}q}{(2\pi)^{4}}\frac{2M_B}{[(P-q)^{2}-m_\phi^{2}+i\epsilon][q^{2}-M_B^{2}+i\epsilon]}.
\end{equation}
This loop function is divergent and needs to be properly regularized. In principle, one can either adopt the dimensional regularization scheme or
the cutoff regularization scheme.  In Ref.~\cite{Altenbuchinger:2013vwa}, a so-called heavy-quark symmetry (HQS) inspired regularization scheme has been suggested, which manifestly satisfies both
the chiral power counting and the heavy-quark spin and flavor symmetry up to $1/M_H$, where $M_H$ is a generic heavy-hadron mass. In the present work, we adopt the HQS regularization scheme, which reads

\begin{equation}
G_{HQS}=G_{\overline{MS}}-\frac{2\mathring{M}}{16\pi^{2}}\left(\log\left(\frac{\mathring{M}^{2}}{\mu^{2}}\right)-2\right)+\frac{2 m_\mathrm{sub}}{16\pi^{2}}\left(\log\left(\frac{\mathring{M}^{2}}{\mu^{2}}\right)+a\right),
\end{equation}
\begin{eqnarray}\label{Eq:MSbar}
   \begin{split}
      G_{\overline{MS}}(s,M^{2},m^{2}) & =\frac{2 M}{16\pi^{2}}\left[\frac{m^{2}-M^{2}+s}{2s}\log\left(\frac{m^{2}}{M^{2}}\right)\right.\\
                                       & -\frac{q}{\sqrt{s}}(\log[2q\sqrt{s}+m^{2}-M^{2}-s]+\log[2q\sqrt{s}-m^{2}+M^{2}-s]\\
                                       & -\log[2q\sqrt{s}+m^{2}-M^{2}+s]-\log[2q\sqrt{s}-m^{2}+M^{2}+s])\\
                                       & \left.+\left(\log\left(\frac{M^{2}}{\mu^{2}}\right)\underline{-2}\right)\right].
   \end{split}
\end{eqnarray}
In the above equations, $m_\mathrm{sub}$ is a generic pseudoscalar meson mass, which can take the value of $m_\pi$ in the $u$, $d$ flavor case or
an average of the pion, the kaon, and the eta masses in the $u$, $d$, and $s$ three-flavor case.  $\mathring{M}$ is the chiral limit value of the charmed or bottom baryon masses. In the present study, we use the averaged antitriplet and sextet charmed or bottom baryon masses given in Table \ref{Table:masses}, instead. The difference is of higher chiral order.
Clearly, the HQS inspired regularization method is a straightforward extension of the minimal subtraction scheme, which, in spirit, is very similar to
the extended-on-mass-shell scheme~\cite{Fuchs:2003qc}. In our present work, for the sake of comparison, we also present results obtained with the cutoff regularization scheme, where
\begin{equation}
G_\mathrm{cut}=\int\limits_0^\Lambda \frac{q^2\,dq}{2\pi^2}\frac{E_M+E_m}{2 E_M  E_m}\frac{2 M}{s-(E_M+E_m)^2+i\epsilon},
\end{equation}
with $E_M=\sqrt{q^2+M^2}$ and $E_m=\sqrt{q^2+m^2}$. In the UChPT framework, one usually replaces the underlined $-2$ of Eq.~(\ref{Eq:MSbar}) by a subtraction constant to  approximate unknown short-range or higher-order interactions. In the following, we  refer to this regularization scheme as the $\overline{\mathrm{MS}}$ scheme.

\begin{figure}[t]
\includegraphics[width=0.45\textwidth]{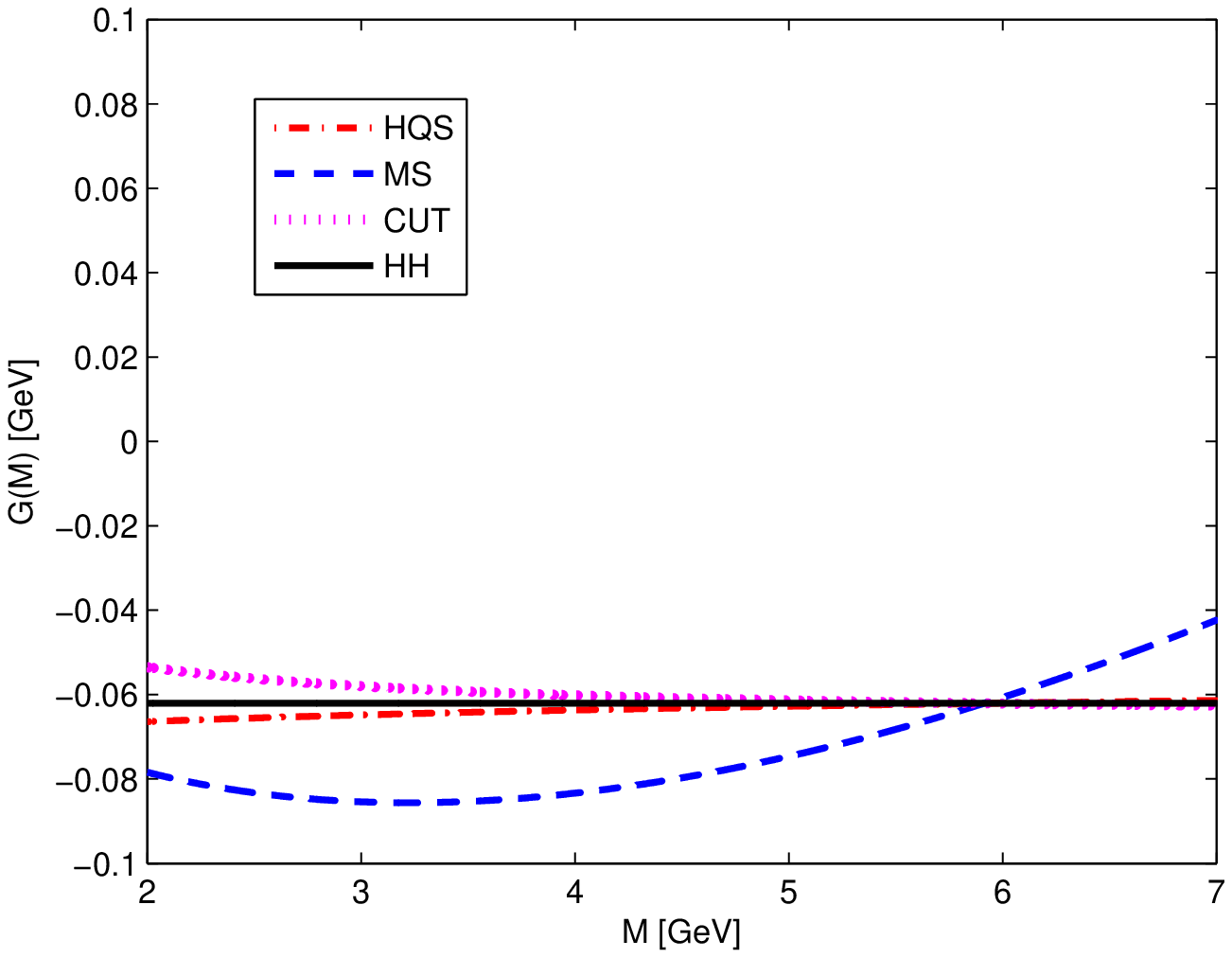}
\includegraphics[width=0.45\textwidth]{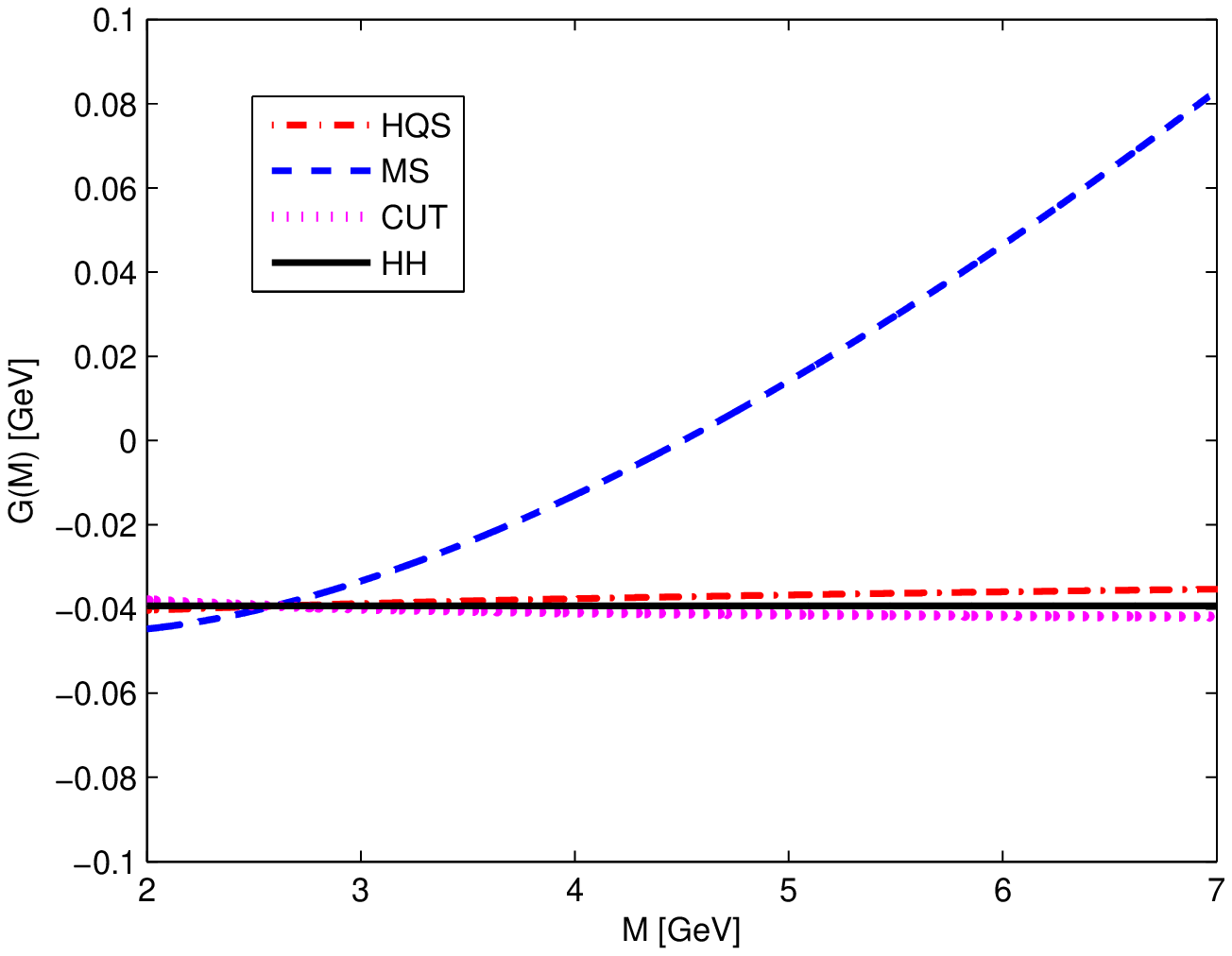}
\caption{ Loop function $G(M)$ as a function of the heavy-hadron mass $M$ in different regularization schemes: HQS,
$\overline{\mathrm{MS}}$ , the cutoff regularization scheme (CUT ), and the exact heavy-quark limit  (HH). The subtraction constants or cutoff values have been fixed by reproducing the
$\Lambda_b(5912)$ (left panel) or the $\Lambda_c(2595)$ (right panel). In calculating the loop function $G$, the pseudoscalar meson mass is fixed at that of the pion $m=138$ MeV, and the renormalization
scale in the dimensional regularization methods is fixed at $\mu=1$ GeV.}
\label{Fig:loop}
\end{figure}

In Fig.~\ref{Fig:loop}, the loop functions $G$ calculated in different regularization schemes are compared with each other. The subtraction constants or cutoff values have been fixed by reproducing the
$\Lambda_b(5912)$ (left panel) or the $\Lambda_c(2595)$ (right panel). In calculating the loop function $G$, the pseudoscalar meson mass is fixed at that of the pion $m=138$ MeV, and the renormalization
scale in the dimensional regularization methods is fixed at $\mu=1$ GeV. The loop function in the exact heavy-quark limit is obtained by replacing  $\mathring{M}$ with $M$ and expanding $G_{HQS}$ in inverse powers of $M$ up to $\mathcal{O}(1/M)$~\cite{Altenbuchinger:2013vwa}. It is clear that the loop functions of both the HQS scheme and the cutoff scheme seem to satisfy the heavy-quark symmetry to a few percent, while
the naive $\overline{\mathrm{MS}}$ scheme strongly  breaks the symmetry, consistent with the finding of Ref.~\cite{Altenbuchinger:2013vwa}.  To be conservative, in the following study of dynamically generated charmed (bottom) baryons, unless otherwise mentioned, we shall present the results obtained in both regularization schemes.

\begin{table*}[t]
     \renewcommand{\arraystretch}{1.6}
     \setlength{\tabcolsep}{0.55cm}
     \centering
     \caption{\label{table:par}Numerical values of isospin and SU3-multiplet averaged masses, the pion decay constant $f_\pi$, and the SU(3) averaged
     pseudoscalar meson decay constant $f_0$ (in units of MeV)~\cite{Agashe:2014kda}. The mass of the $\Omega^*_b$ is taken from Ref.~\cite{AliKhan:1999yb}.\label{Table:masses} }
     \begin{tabular}{cccccccc}
     \hline\hline
    $\mathring{M}_c^{[\bar{3}]}$  &  $M_{\Lambda_c}$&$M_{\Xi_c}$ & $\mathring{M}_c^{[6]} $ & $M_{\Sigma_c}$  & $M_{\Xi'_c}$   & $M_{\Omega_c}$       \\
   $2408.5$    & 2286.5& 2469.5 & 2534.9 & 2453.5  & 2576.8& 2695.2\\   \hline
  $\mathring{M}_b^{[\bar{3}]}$  &  $M_{\Lambda_b}$& $M_{\Xi_b}$  &$\mathring{M}_b^{[6]}$ & $M_{\Sigma_b}$ & $M_{\Xi'_b}$   & $M_{\Omega_b}$    \\
   $5732.8$    & 5619.4 &   5789.5 &5890.0&5813.4&   5926& 6048\\ \hline
   $M_{\Sigma^{*}_c}$ & $M_{\Xi^{*}_c}$ & $M_{\Omega^{*}_c}$ & $\mathring{M}_{c^{*}}^{[6]}$ & $M_{\Sigma^{*}_b}$ & $M_{\Xi^{*}_b}$ & $M_{\Omega^{*}_b}$ & \\
   2517.9 & 2645.9 & 2765.9 & 2601.9 & 5833.5 & 5949.3 & 6069 & \\ \hline
   $\mathring{M}_{b^{*}}^{[6]}$ & $m_\pi$ & $m_K$  &    $m_\eta$ & $m_{sub}$ & $f_\pi$ & $f_0=1.17f_\pi$ \\
   5911.35 & 138.0  & 495.6 & 547.9 & 368.1 &92.21 & 107.8\\
 \hline\hline
    \end{tabular} 
\end{table*}

\section{Results and discussions}
At leading order, the only unknown parameter in the UChPT  is related to the regularization of the loop function $G$, i.e.,
the subtraction constant $a$ in the dimensional regularization scheme or the cutoff value $\Lambda$ in the cutoff regularization scheme.  Conventionally, in the latter method one
often chooses a cutoff of the order of 1 GeV (the chiral symmetry breaking scale). Requiring the $G$ function evaluated at threshold to be equal in both methods,
one can fix a ``natural'' value for the subtraction constant. In most cases, the above-mentioned prescription allows one to assign some of the dynamically generated states
to their experimental counterparts. Once the identification is done, one can slightly fine-tune $\Lambda$ or $a$ so that the dynamically generated state coincides with its experimental counterpart and
then use the so-obtained $\Lambda$ or $a$ to make predictions. We follow the same line of argument in the present work. As in previous works, we can identify the $\Lambda_c(2595)$ and the
$\Lambda_b(5912)$ as dynamically generated states in their respective coupled channels.

Approximate heavy-quark spin symmetry implies that the interactions between a ground-state spin-1/2 (3/2) baryon and a pseudoscalar meson are the same in the limit of infinite heavy-quark masses. Therefore, one can extend the LO study of the $1/2^-$ sector to the $3/2^-$ sector.
 As a first approximation, we only need to replace the masses of the  $1/2^+$ baryons by their $3/2^+$ counterparts (see Table \ref{Table:masses}). Searching for poles
on the complex plane,  we find two states in the charmed and bottom sector with the same quantum numbers as those of the $\Lambda^{*}_{c}(2625)$ and $\Lambda^{*}_{b}(5920)$.  Namely,
they can be identified as the heavy-quark spin partners of the $\Lambda_c(2595)$ and $\Lambda_b(5912)$.

 It should be noted that, unlike the heavy meson sector, the present lattice QCD simulations of charmed~\cite{  Briceno:2012wt,Alexandrou:2014sha,Liu:2009jc,Namekawa:2013vu,Brown:2014ena} or bottom~\cite{Liu:2009jc,Lewis:2008fu,Brown:2014ena} baryons still focus on the ground states
 with the exception of Refs.~\cite{Padmanath:2013zfa} and  \cite{Meinel:2012qz}, where excited triply charmed and bottom states were studied, respectively.
Future lattice QCD simulations of the excited singly charmed and bottom baryons will be extremely valuable to test the predictions of the present work and those of other studies.\footnote{We note that preliminary results on the excited-state spectroscopy of singly and doubly charmed baryons  have recently been presented at conferences~\cite{  Padmanath:2013bla,Padmanath:2013laa}.}

\begin{table}[t]
      \centering
      \caption{Dynamically generated bottom baryons of $J^P=1/2^-$.
       The subtraction constant is fixed in a way such that the $\Lambda_{b}(5912)$ mass is produced to be 5912 MeV with $a=-14.15$. All energies are in units of MeV and $(S,I)^M$ denotes (strangeness, isospin)$^\mathrm{SU(3) multiplet}$.}\label{Table:bsub}
      \begin{tabular}{lclll}
         \hline\hline
         Pole position &  $(S,I)^{M}$ & Main channels (threshold)&  Exp.~\cite{Agashe:2014kda}\\\hline
         $(6081,-i57)$ &  $(0,1)^{[\overline{3}]}$ & $\Xi_{b}K$(6285.1) &  \\
         $(5921,0)$ &  $(0,0)^{[\overline{3}]}$ & $\Xi_{b}K$(6285.1) &  \\
         $(5868,0)$ &  $(-1,\frac{1}{2})^{[\overline{3}]}$ & $\Lambda_{b}\overline{K}$(6115.0) &  \\
         $(6118,-i50)$ &  $(-1,\frac{1}{2})^{[\overline{3}]}$ & $\Xi_{b}\eta$(6337.4),$\Xi_{b}\pi$(5927.5) & \\
         $(6201,0)$ &  $(-2,0)^{[\overline{3}]}$ & $\Xi_{b}\overline{K}$(6285.1)  & \\
         \hline

         $(6201,0)$ &  $(1,\frac{1}{2})^{[6]}$ & $\Sigma_{b}K$(6308.6) &  \\
         $(5967,-i9)$ &  $(0,1)^{[6]}$ & $\Sigma_{b}\pi$(5951.0)  & \\
         $(6223,-i14)$ &  $(0,1)^{[6]}$ & $\Sigma_{b}\eta$(6360.9) &  \\
         $(5912,0)$ &  $(0,0)^{[6]}$ & $\Sigma_{b}\pi$(5951.0) &  $\Lambda_b(5912)$\\
         $(6307,-i12)$ &  $(0,0)^{[6]}$ & $\Xi'_{b}K$(6421.6)  & \\
         $(6213,-i25)$ & $( -1,\frac{3}{2})^{[6]}$ & $\Sigma_{b}\overline{K}$(6308.6)  & \\
         $(5955,0)$ & $( -1,\frac{1}{2})^{[6]}$ & $\Sigma_{b}\overline{K}$(6308.6)  & \\
         $(6101,-i15)$ & $( -1,\frac{1}{2})^{[6]}$ & $\Xi'_{b}\pi$(6064.0)  & \\
         $(6364,-i27)$ &  $(-1,\frac{1}{2})^{[6]}$ & $\Omega_{b}K$(6543.6) & \\
         $(6361,-i59)$ &  $(-2,1)^{[6]}$ & $\Omega_{b}\pi$(6186.0)  & \\
         $(6169,0)$ &  $(-2,0)^{[6]}$ & $\Xi'_{b}\overline{K}$(6421.6),$\Omega_{b}\eta$(6595.6)  & \\
         \hline\hline
      \end{tabular}
\end{table}

\begin{table}
      \centering
      \caption{The same as Table \ref{Table:bsub}, but obtained in  the cutoff regularization scheme with $\Lambda=2.17$ GeV.}\label{Table:bcut}
      \begin{tabular}{lclll}
         \hline\hline
         Pole positions & $(S,I)^{M}$ & Main channels &  Ref.~\cite{Agashe:2014kda}\\
         \hline
         $(6083,-i72)$ &  $(0,1)^{[\overline{3}]}$ & $\Xi_{b}K$(6285.1) & \\
         $(5908,0)$ &  $(0,0)^{[\overline{3}]}$ & $\Xi_{b}K$(6285.1)  & \\
         $(5867,0)$ &  $(-1,\frac{1}{2})^{[\overline{3}]}$ & $\Lambda_{b}\overline{K}$(6115.0) &  \\
         $(6116,-i55)$ &  $(-1,\frac{1}{2})^{[\overline{3}]}$ & $\Xi_{b}\eta$(6337.4),$\Xi_{b}\pi$(5927.5)&  \\
         $(6198,0)$ &  $(-2,0)^{[\overline{3}]}$ & $\Xi_{b}\overline{K}$(6285.1) &  \\\hline

         $(6221,0)$ & $(1,\frac{1}{2})^{[6]}$ & $\Sigma_{b}K$(6308.6) &  \\
         $(5966,-i9)$ &  $(0,1)^{[6]}$ & $\Sigma_{b}\pi$(5951.0) &  \\
         $(6234,-i20)$ &  $(0,1)^{[6]}$ & $\Sigma_{b}\eta$(6360.9) &  \\
         $(5912,0)$ &  $(0,0)^{[6]}$ & $\Sigma_{b}\pi$(5951)  &  $\Lambda_b(5912)$\\
         $(6305,-i17)$ &  $(0,0)^{[6]}$ & $\Xi'_{b}K$(6421.6) & \\
         $(6226,-i27)$ & $( -1,\frac{3}{2})^{[6]}$ & $\Sigma_{b}\overline{K}$(6308.6) &  \\
         $(5951,0)$ & $( -1,\frac{1}{2})^{[6]}$ & $\Sigma_{b}\overline{K}$(6308.6) &  \\
         $(6089,-i10)$ & $( -1,\frac{1}{2})^{[6]}$ & $\Xi'_{b}\pi$(6064.0)  & \\
         $(6343,-i35)$ &  $(-1,\frac{1}{2})^{[6]}$ & $\Omega_{b}K$(6543.6) & \\
         $(6347,-i55)$ &  $(-2,1)^{[6]}$ & $\Omega_{b}\pi$(6186.0) &  \\
         $(6139,0)$ &  $(-2,0)^{[6]}$ & $\Xi'_{b}\overline{K}$(6421.6)& \\
         \hline\hline
      \end{tabular}
\end{table}

\begin{table}[!htb]
      \centering
      \caption{Dynamically generated bottom baryons of $J^P=3/2^-$.
       The subtraction constant is fixed in a way such that the $\Lambda^*_{b}(5920)$ mass is produced to be 5920 MeV with $a=-16.27$. All energies are in units of MeV and $(S,I)^M$ denotes (strangeness, isospin)$^\mathrm{SU(3) multiplet}$.}\label{Table:bsub2}
      \begin{tabular}{lclll}
         \hline\hline
         Pole position &  $(S,I)^{M}$ & Main channels (threshold)&  Ref.~\cite{Agashe:2014kda}\\\hline
         $(6181,0)$ &  $(1,\frac{1}{2})^{[6]}$ & $\Sigma^{*}_{b}K$(6329.1) & \\
         $(5971,0)$ &  $(0,1)^{[6]}$ & $\Sigma^{*}_{b}\pi$(5971.5) &  \\
         $(6202,-i12)$ &  $(0,1)^{[6]}$ & $\Sigma^{*}_{b}\eta$(6381.4) &  \\
         $(5920,0)$ &  $(0,0)^{[6]}$ & $\Sigma^{*}_{b}\pi$(5971.5) &   $\Lambda^{*}_b(5920)$\\
         $(6289,-i11)$ &  $(0,0)^{[6]}$ & $\Xi^{*}_{b}K$(6444.9) &  \\
         $(6197,-i19)$ & $( -1,\frac{3}{2})^{[6]}$ & $\Sigma^{*}_{b}\overline{K}$(6329.1) &  \\
         $(5950,0)$ & $( -1,\frac{1}{2})^{[6]}$ & $\Sigma^{*}_{b}\overline{K}$(6329.1) &  \\
         $(6102,-i7)$ & $( -1,\frac{1}{2})^{[6]}$ & $\Xi^{*}_{b}\pi$(6087.3) &  \\
         $(6344,-i23)$ &  $(-1,\frac{1}{2})^{[6]}$ & $\Omega^{*}_{b}K$(6564.6)&  \\
         $(6349,-i45)$ &  $(-2,1)^{[6]}$ & $\Omega^{*}_{b}\pi$(6207.0) &  \\
         $(6152,0)$ &  $(-2,0)^{[6]}$ & $\Xi^{*}_{b}\overline{K}$(6444.9) &  \\
         \hline\hline
      \end{tabular}
\end{table}

\begin{table}
      \centering
      \caption{The same as Table \ref{Table:bsub}, but obtained in  the cutoff regularization scheme with $\Lambda=2.60$ GeV.}\label{Table:bcut2}
      \begin{tabular}{lclll}
         \hline\hline
         Pole position &  $(S,I)^{M}$ & Main channels (threshold)& Ref.~\cite{Agashe:2014kda}\\\hline
         $(6193,0)$ &  $(1,\frac{1}{2})^{[6]}$ & $\Sigma^{*}_{b}K$(6329.1) &  \\
         $(5971,0)$ &  $(0,1)^{[6]}$ & $\Sigma^{*}_{b}\pi$(5971.5) &  \\
         $(6205,-i15)$ &  $(0,1)^{[6]}$ & $\Sigma^{*}_{b}\eta$(6381.4) &  \\
         $(5920,0)$ &  $(0,0)^{[6]}$ & $\Sigma^{*}_{b}\pi$(5971.5) &   $\Lambda^{*}_b(5920)$\\
         $(6281,-i13)$ &  $(0,0)^{[6]}$ & $\Xi^{*}_{b}K$(6444.9) &  \\
         $(6202,-i19)$ & $( -1,\frac{3}{2})^{[6]}$ & $\Sigma^{*}_{b}\overline{K}$(6329.1) &  \\
         $(5946,0)$ & $( -1,\frac{1}{2})^{[6]}$ & $\Sigma^{*}_{b}\overline{K}$(6329.1) &  \\
         $(6091,-i3)$ & $( -1,\frac{1}{2})^{[6]}$ & $\Xi^{*}_{b}\pi$(6087.3) &  \\
         $(6321,-i24)$ &  $(-1,\frac{1}{2})^{[6]}$ & $\Omega^{*}_{b}K$(6564.6) & \\
         $(6331,-i39)$ &  $(-2,1)^{[6]}$ & $\Omega^{*}_{b}\pi$(6207.0) &  \\
         $(6128,0)$ &  $(-2,0)^{[6]}$ & $\Xi^{*}_{b}\overline{K}$(6444.9) &  \\
         \hline\hline
      \end{tabular}
\end{table}

\subsection{Dynamically generated bottom baryons}

In Refs.~\cite{GarciaRecio:2012db,Liang:2014eba}, the $\Lambda_b(5912)$ is found to be dynamically generated. In Ref.~\cite{GarciaRecio:2012db}, the dominant coupled channel
is identified as $\bar{B}N$, while in Ref.~\cite{Liang:2014eba} it is identified as $\bar{B}^* N$.  In our approach, this state appears naturally as a
$\Sigma_b \pi$ state.  It is useful to point out the major differences among the approaches of Ref.~\cite{GarciaRecio:2012db}, Ref.~\cite{Liang:2014eba}, and the present work.
The kernel looks similar in all the three cases. However, the three approaches differ in the number of coupled channels included and how the transition amplitudes between
different channels are obtained. In Ref.~\cite{GarciaRecio:2012db}, the transition amplitudes between different coupled channels are obtained by invoking the
SU(6) symmetry and heavy-quark spin symmetry, while in Ref.~\cite{ Liang:2014eba}, they are obtained through the vector meson exchange or pion exchange.
The number of coupled channels considered is the largest in Ref.~\cite{GarciaRecio:2012db}, while it is the smallest in our approach. In other words, we only consider the
minimum number of channels needed to construct the LO chiral Lagrangians. In addition, up to the order at which we are working, the [$\bar{3}$] multiplet and the [$6$] multiplet do not mix.  Therefore, one needs to be careful when
comparing our predictions with those of Refs.~\cite{GarciaRecio:2012db,Liang:2014eba}.

To study the interaction between the ground-state bottom baryons and the pseudoscalar mesons, we fix the cutoff value or the subtraction constant
in such a way that the mass of the $\Lambda_b(5912)$ is produced to be 5912 MeV.
Broken SU(3) chiral symmetry then predicts a number of additional resonances or bound states as shown in Tables \ref{Table:bsub} and \ref{Table:bcut}. It can be seen
that in addition to the $\Lambda_b(5912)$, both regularization schemes generate a number of other states, the experimental counterparts of which cannot be  identified .
Future experiments are strongly encouraged to search for these states.

In Ref.~\cite{GarciaRecio:2012db}, only $(S,I)=(0,0)$ and $(-1,1/2)$ sectors are studied. For $J=1/2$, three $\Lambda_b$ states and three $\Xi_b$ states are identified.   From the couplings of those dynamically generated
states to the corresponding coupled channels (Tables III and IV of Ref.~\cite{GarciaRecio:2012db}), it is clear that none of those states couples dominantly to the coupled channels  considered in the present study.
For instance,  their $\Lambda_b(5912)$ and the bound state with $M=6009.3$ MeV couple only moderately to $\Sigma_b\pi$ and $\Lambda_b\eta$, respectively.  The same is true for the two $\Xi_b$ states with
$M=6035.4$ MeV and $M=6072.8$ MeV.

In Ref.~\cite{Liang:2014eba}, ten states are found in the $J^P=1/2^-$ sector. Among them, one $I=0$ state with $M=5969.5$  and one $I=1$ state with $M=6002.8$ MeV  couple strongly
to $\Sigma_b\pi$.  Because of the different coupled channels considered in both works and the fact that the [$\bar{3}$] and [6] multiplets do not mix at leading order chiral perturbation theory, we must refuse the temptation to associate them to our dynamically generated states.
One needs to keep in mind that in our present work only the smallest number of coupled channels that are dictated by approximate SU(3) chiral symmetry is taken into account. The introduction of additional coupled channels inevitably requires further less justified assumptions.

In principle, one can use the same subtraction constants or cutoff values for the $3/2^-$ sector. However, to make the prediction more precise, we slightly fine-tune them
to reproduce the experimental mass of the $\Lambda_b^*(5920)$. The relative change of the subtraction constant or cutoff value reflects the effect of the heavy-quark symmetry breaking (at least in our framework). The results are shown in Tables ~\ref{Table:bsub2} and \ref{Table:bcut2},  For the reason we mentioned above, the results are quite similar to those in the $J^P=1/2^-$ sector. In total, as
in the $1/2^-$ sector,  11 states are found, the $J^P=1/2^-$ partners of which can be easily identified.

\begin{table}[!htb]
      \centering
      \caption{Dynamically generated charmed baryons of $J^P=1/2^-$.
        The subtraction constant is fixed in a way such that the $\Lambda_{c}(2595)$ mass is produced to be 2591 MeV with $a=-8.27$.  All energies are in units of MeV and $(S,I)^M$ denotes (strangeness, isospin)$^\mathrm{SU(3) multiplet}$.}\label{Table:csub}
      \begin{tabular}{lclll}
         \hline\hline
         Pole position &
          $(S,I)^{M}$ & Main channels  (threshold) & Ref.~\cite{Agashe:2014kda} \\\hline
         $(2721,0)$ &  $(0,0)^{[\overline{3}]}$ & $\Xi_{c}K$(2965.1) & $\Lambda_c(2765)?$  \\
         $(2623,-i12)$ &  $(-1,\frac{1}{2})^{[\overline{3}]}$ & $\Lambda_{c}\overline{K}$(2782.1) &  \\
         $(2965,0)$ &  $(-2,0)^{[\overline{3}]}$ & $\Xi_{c}\overline{K}$(2965.1) &  \\\hline

         $(2948,0)$ &  $(1,\frac{1}{2})^{[6]}$ & $\Sigma_{c}K$(2949.1) &  \\
         $(2674,-i51)$ &  $(0,1)^{[6]}$ & $\Sigma_{c}\pi$(2591.5) &  \\
         $(2999,-i16)$ & $(0,1)^{[6]}$ & $\Sigma_{c}\eta$(3001.4),$\Xi'_{c}K$(3072.4)  &  \\
         $(2591,0)$ &  $(0,0)^{[6]}$ & $\Sigma_{c}\pi$(2591.5)  & $\Lambda_c(2595)$  \\
         $(3069,-i12)$ &  $(0,0)^{[6]}$ & $\Xi'_{c}K$(3072.4)    & $\Lambda_c(2940)?$\\
         $\underline{(2947,-i34)}$ &  $( -1,\frac{3}{2})^{[6]}$ & $\Sigma_{c}\overline{K}$(2949.1) &  \\
         $(2695,0)$ &  $(-1,\frac{1}{2})^{[6]}$ & $\Sigma_{c}\overline{K}$(2949.1) & \\
         $(2827,-i55)$ &  $(-1,\frac{1}{2})^{[6]}$ & $\Xi'_{c}\pi$(2714.7)  & $\Xi_c(2790)$?\\
         $\underline{(3123,-i44)}$ &  $(-1,\frac{1}{2})^{[6]}$ & $\Omega_{c}K$(3190.8)  & $\Xi_c(3123)?$\\
         $(2946,0)$ &  $(-2,0)^{[6]}$ & $\Xi'_{c}\overline{K}$(3072.4),$\Omega_{c}\eta$(3243.1) &  \\
         \hline\hline
      \end{tabular}
\end{table}
\begin{table}
      \centering
      \caption{The same as Table \ref{Table:csub}, but obtained in the cutoff regularization scheme with $\Lambda=1.35$ GeV.}\label{Table:ccut}
      \begin{tabular}{lclll}
         \hline\hline
         Pole positions&  $(S,I)^{M}$ & Main channels (threshold) & Ref.~\cite{Agashe:2014kda} \\\hline
         $(2707,0)$ &  $(0,0)^{[\overline{3}]}$ & $\Xi_{c}K$(2965.1) & $\Lambda_c(2765)?$\\
         $(2622,-i12)$ &  $(-1,\frac{1}{2})^{[\overline{3}]}$ & $\Lambda_{c}\overline{K}$(2782.1) &  \\
         $\underline{(2965,0)}$ &  $(-2,0)^{[\overline{3}]}$ & $\Xi_{c}\overline{K}$(2965.1) &  \\\hline

         $\underline{(2949,0)}$ &  $(1,\frac{1}{2})^{[6]}$ & $\Sigma_{c}K$(2949.1) &  \\
         $(2672,-i53)$ &  $(0,1)^{[6]}$ & $\Sigma_{c}\pi$(2591.5) &  \\
         $\underline{(2996,-i21)}$ & $( 0,1)^{[6]}$ & $\Sigma_{c}\eta$(3001.4)  &  \\
         $(2591,0)$ &  $(0,0)^{[6]}$ & $\Sigma_{c}\pi$(2591.5)   & $\Lambda_c(2595)$\\
         $(3072,-i15)$ &  $(0,0)^{[6]}$ & $\Xi'_{c}K$(3072.4) &  $\Lambda_c(2940)?$\\
         $\underline{(2946,-i35)}$ &  $( -1,\frac{3}{2})^{[6]}$ & $\Sigma_{c}\overline{K}$(2949.1) &  \\
         $(2683,0)$ &  $(-1,\frac{1}{2})^{[6]}$ & $\Sigma_c\overline{K}$(2949.1)&  \\
         $(2813,-i44)$ &  $(-1,\frac{1}{2})^{[6]}$ & $\Xi'_{c}\pi$(2714.7)  & $\Xi_c(2790)$?\\
         $(3121,-i61)$ &  $(-1,\frac{1}{2})^{[6]}$ & $\Omega_cK$(3190.8) &$\Xi_c(3123)?$ \\
         $(2909,0)$ &  $(-2,0)^{[6]}$ & $\Xi'_{c}K$(3072.4) &  \\
         \hline\hline
      \end{tabular}
\end{table}

\begin{table}[!htb]
      \centering
      \caption{Dynamically generated charmed baryons of $J^P=3/2^-$.
        The subtraction constant is fixed in a way such that the $\Lambda^*_{c}(2625)$ mass is produced to be 2625 MeV with $a=-12.0$.  All energies are in units of MeV and $(S,I)^M$ denotes (strangeness, isospin)$^\mathrm{SU(3) multiplet}$.}\label{Table:csub2}
      \begin{tabular}{lclll}
         \hline\hline
         Pole position &
          $(S,I)^{M}$ & Main channels  (threshold) & Exp.~\cite{Agashe:2014kda} \\\hline
         $(2952,0)$ &  $(1,\frac{1}{2})^{[6]}$ & $\Sigma^{*}_{c}K$(3013.5) &  \\
         $(2685,-i15)$ &  $(0,1)^{[6]}$ & $\Sigma^{*}_{c}\pi$(2655.9) & \\
         $(2977,-i23)$ & $(0,1)^{[6]}$ & $\Sigma^{*}_{c}\eta$(3065.8)  & \\
         $(2625,0)$ &  $(0,0)^{[6]}$ & $\Sigma^{*}_{c}\pi$(2655.9) & $\Lambda^{*}_c(2625)$  \\
         $(3066,-i19)$ &  $(0,0)^{[6]}$ & $\Xi^{*}_{c}K$(3141.5)   &   \\
         $(2968,-i33)$ &  $( -1,\frac{3}{2})^{[6]}$ & $\Sigma^{*}_{c}\overline{K}$(3013.5) &  \\
         $(2656,0)$ &  $(-1,\frac{1}{2})^{[6]}$ & $\Sigma^{*}_{c}\overline{K}$(3013.5)&  \\
         $(2827,-i17)$ &  $(-1,\frac{1}{2})^{[6]}$ & $\Xi^{*}_{c}\pi$(2783.9)  &  $\Xi_c^*(2815)?$   \\
         $(3113,-i45)$ &  $(-1,\frac{1}{2})^{[6]}$ & $\Omega^{*}_{c}K$(3261.5)& \\
         $(3118,-i80)$ &  $(-2,1)^{[6]}$ & $\Omega^{*}_{c}\pi$(2903.9) &  \\
         $(2885,0)$ &  $(-2,0)^{[6]}$ & $\Xi^{*}_{c}\overline{K}$(3141.5) &  \\
         \hline\hline
      \end{tabular}
\end{table}
\begin{table}
      \centering
      \caption{The same as Table \ref{Table:csub}, but obtained in the cutoff regularization scheme with $\Lambda=2.13$ GeV.}\label{Table:ccut2}
      \begin{tabular}{lclll}
         \hline\hline
         Pole position &
          $(S,I)^{M}$ & Main channels  (threshold) & Exp.~\cite{Agashe:2014kda} \\\hline
         $(2962,0)$ &  $(1,\frac{1}{2})^{[6]}$ & $\Sigma^{*}_{c}K$(3013.5) &  \\
         $(2684,-i15)$ &  $(0,1)^{[6]}$ & $\Sigma^{*}_{c}\pi$(2655.9) &  \\
         $(2980,-i28)$ & $(0,1)^{[6]}$ & $\Sigma^{*}_{c}\eta$(3065.8)  &  \\
         $(2625,0)$ &  $(0,0)^{[6]}$ & $\Sigma^{*}_{c}\pi$(2655.9)  & $\Lambda^{*}_c(2625)$  \\
         $(3059,-i22)$ &  $(0,0)^{[6]}$ & $\Xi^{*}_{c}K$(3141.5)   &   \\
         $(2974,-i33)$ &  $( -1,\frac{3}{2})^{[6]}$ & $\Sigma^{*}_{c}\overline{K}$(3013.5) &  \\
         $(2653,0)$ &  $(-1,\frac{1}{2})^{[6]}$ & $\Sigma^{*}_{c}\overline{K}$(3013.5) & \\
         $(2816,-i13)$ &  $(-1,\frac{1}{2})^{[6]}$ & $\Xi^{*}_{c}\pi$(2783.9)  &   $\Xi_c^*(2815)?$  \\
         $(3093,-i51)$ &  $(-1,\frac{1}{2})^{[6]}$ & $\Omega^{*}_{c}K$(3261.5) &  \\
         $(3103,-i74)$ &  $(-2,1)^{[6]}$ & $\Omega^{*}_{c}\pi$(2903.9) &  \\
         $(2858,0)$ &  $(-2,0)^{[6]}$ & $\Xi^{*}_{c}\overline{K}$(3141.5) &  \\
         \hline\hline
      \end{tabular}
\end{table}

\subsection{Dynamically generated charmed baryons}

Once the subtraction constant is fixed in the HQS approach, one can use the same constant to predict the counterparts of the dynamically generated bottom baryons.
We have performed such a calculation and found that the $\Lambda_c(2595)$ can indeed be identified as a $\Sigma_c \pi$ state, as first pointed out
in Refs.~\cite{Lutz:2003jw,Hofmann:2005sw}. To account for moderate heavy-quark flavor symmetry breaking corrections, we slightly fine-tune the subtraction constant in the dimensional regularization scheme or
the cutoff value in the cutoff regularization scheme so that the mass of the $\Lambda_c(2595)$ is reproduced to be 2591 MeV.\footnote{Experimentally, the
$\Lambda_c(2595)$ is found at $2592.25\pm0.28$ MeV with a width of $2.6\pm0.6$ MeV~\cite{Agashe:2014kda} . We need to slightly increase $f_0$ to put
the $\Lambda_c(2595)$ exactly at this position because of the closeness of the $\Sigma_c\pi$ threshold.}  The predictions are then tabulated in Tables \ref{Table:csub} and \ref{Table:ccut}.

A comparison with the predictions of Refs.~\cite{GarciaRecio:2008dp,Liang:2014kra} is again complicated by the same factors as mentioned previously.  For instance,
four $(S=0,I=0)$ states and five $(S=0,I=1)$ states are predicted in Ref.~\cite{Liang:2014kra}. The number of dynamically generated states in Ref.~\cite{GarciaRecio:2008dp} is even larger.
Somehow, it seems that the number of states generated is proportional to the number of coupled channels considered.

In addition, our $\Lambda_c(2595)$ is predominantly a $\Sigma_c \pi$ state, where it is more of a $DN$ state in Ref.~\cite{Liang:2014kra} and a $D^* N$ state in Ref.~\cite{GarciaRecio:2008dp}.
Despite of the different dominant components, it is clear that coupled channel effects or multiquark components may not be negligible in the wave function of the $\Lambda_c(2595)$. The same can
be said about the $\Lambda_b(5912$).

In Tables \ref{Table:csub}, we have temporarily identified the states appearing at $\sqrt{s}=(2721,0)$ MeV, $\sqrt{s}=(3069-i12)$ MeV, $\sqrt{s}=(2827-i55)$ MeV, and $\sqrt{s}=(3123,-i44)$ MeV as  the $\Lambda_c(2765)$, $\Lambda_c(2940)$, $\Xi_c(2790)$, and $\Xi_c(3123)$. These identifications are mainly
based on the masses of these states~\cite{Agashe:2014kda}. Since the spin parities of  these states are not yet known, the associations of our states with their experimental counterparts should be taken with care. A second complication comes from the fact that coupled channels other than those considered here may not be negligible as can be seen from Fig.~\ref{Fig:threshold}.

In Ref.~\cite{He:2006is}, the $\Lambda_c(2940)$ was suggested to be a molecular state with spin parity $J^P=1/2^-$ or $3/2^-$ because of its proximity to the $D^{*0} p$ threshold.  In Ref.~\cite{GarciaRecio:2008dp}, none of the dynamically generated states with $J^P=1/2^-$ or $3/2^-$ can be associated to the $\Lambda(2940)$. In Ref.~\cite{Liang:2014kra}, a state at $2959$ MeV with a small width could be associated to the $\Lambda_c(2940)$, which, however, couples mostly $\rho \Sigma_c$.  In our present study, since the $DN$ ($D^*N$) channels are not taken into account explicitly, we have found only two states located about 3050 MeV (see Tables ~\ref{Table:csub} and \ref{Table:csub2}), one of which we tentatively associate with the $\Lambda_c^{(*)}(2940)$. However, one definitely needs to take into account the missing $D^{(*)} N$ channels to be more conclusive.  It should be noted that in the molecular picture  Dong \textit{et al}. have studied the strong two-body decays of the $\Lambda_c(2940)$ and shown that the $J^P=1/2^+$  assignment is favored~\cite{Dong:2009tg}. Assuming this particular quantum number, they later studied the radiative~\cite{Dong:2010xv} and strong three-body~\cite{Dong:2011ys} decays of the $\Lambda_c(2940)$. The molecular nature of the $\Lambda_c(2940)$ has recently been studied in  the framework of QCD sum rules~\cite{Zhang:2012jk}, the constituent quark model~\cite{Ortega:2012cx}, and the effective Lagrangians method~\cite{He:2010zq}, as well.

In Tables \ref{Table:csub2} and \ref{Table:ccut2}, we tabulate the dynamically generated states in the $3/2^-$ sector.  It should be noted that, compared to the $1/2^-$ sector, an extra pole is produced in the
$(S,I)=(-2,1)$ channel.  On the other hand, its counterpart is found in both the $3/2^-$ and $1/2^-$ bottom sectors. This seems to indicate that the breaking of the heavy-quark flavor symmetry is larger than that of the heavy-quark spin symmetry, as naively expected.

It should be noted that to confirm the identification of the dynamically generated states with their experimental counterparts, one needs to study their decay branching ratios, since many approaches used the masses of these states to fix (some of) their parameters.  Strong and radiative decays are both very important in this respect since they may probe different regions of their wave functions. In the past few years, many such studies of the decays of charmed baryons have been performed, see, e.g., Refs.~\cite{Dong:2009tg,Dong:2010xv,Dong:2011ys,Cheng:2006dk,Chen:2007xf,Zhong:2007gp,Liu:2012sj,Yasui:2014cwa,Gamermann:2010ga}.\footnote{For similar studies in the heavy-flavor mesonic sector; see, e.g., Refs.~\cite{ Gamermann:2007bm,Lutz:2007sk,Dong:2008gb, Liu:2009zg, Ma:2010xx,Nielsen:2010ij,Aceti:2012cb,Dong:2013iqa}.}

\subsection{Further discussions}

Superficially, exact heavy-quark flavor symmetry would dictate that the number of dynamically generated states in the bottom sector and that in the charm sector is the same. A comparison of Tables \ref{Table:bsub} and \ref{Table:csub} (or Tables
\ref{Table:bcut} and \ref{Table:ccut}) shows that this is almost the case, but not exactly.  For instance,  some counterparts of the dynamically generated bottom baryons in the charm sector are missing, such as the
counterparts of the [$\bar{3}$] states at $\sqrt{s}=(6081-i57)$ MeV and $\sqrt{s}=(6118-i50)$ MeV. A closer look at these channels reveals that they simply become too broad and develop a width of $200\sim300$ MeV.  It should be noted that
we have not considered any states broader than 200 MeV in our
study,

The broadening of these states can be traced back partially to the weakening of the corresponding potentials and partially to the calibration of our framework
to reproduce the $\Lambda_b(5912)$ in the bottom sector and to reproduce the $\Lambda_c(2595)$ in the charmed sector. Since the $\Lambda_c(2595)$ is much closer to the threshold of its main coupled channel than
the $\Lambda_b(5912)$, the calibration implies a weaker potential in the charm sector than in the bottom sector. Because of this weakening, the dynamical generation of some charmed baryons requires
a slight readjustment of the potential by changing either $f_0$ or $a$ slightly within a few percent. Otherwise, they will show up as cusps. The pole positions of these states have been underlined to denote such a
fine-tuning.

One should note that we have used an averaged pseudoscalar decay constant, $f_0=1.17 f_\pi$, in our calculations. Using the pion decay constant, $f_0=f_\pi$, will not change qualitatively our results and conclusions but can shift
the predicted baryon masses by a few tens of MeV depending on the particular channel. We have not given explicitly such uncertainties in Tables ~\ref{Table:bsub}, \ref{Table:bcut}, \ref{Table:bsub2}, \ref{Table:bcut2}, \ref{Table:csub},  \ref{Table:ccut}, \ref{Table:csub2}, and \ref{Table:ccut2}, but one should
keep in mind the existence of such uncertainties (or freedom) in our approach. In addition, the differences between the results obtained in the dimensional regularization scheme and those obtained in the cutoff regularization scheme also indicate
inherent theoretical uncertainties of the UChPT method, which can be as large as 30--40 MeV depending on the channel. It should be mentioned that, although formally the dimensional regularization scheme might be
preferred to the cutoff regularization scheme, they yield quite similar results in our present work, both in terms of heavy-quark symmetry conservation and in terms of the prediction of dynamically generated states once the
relevant parameters are fixed in such a way that the $\Lambda_c(2595)$ and the $\Lambda_b(5912)$ are produced.

\begin{figure}[t]
\includegraphics[width=0.48\textwidth]{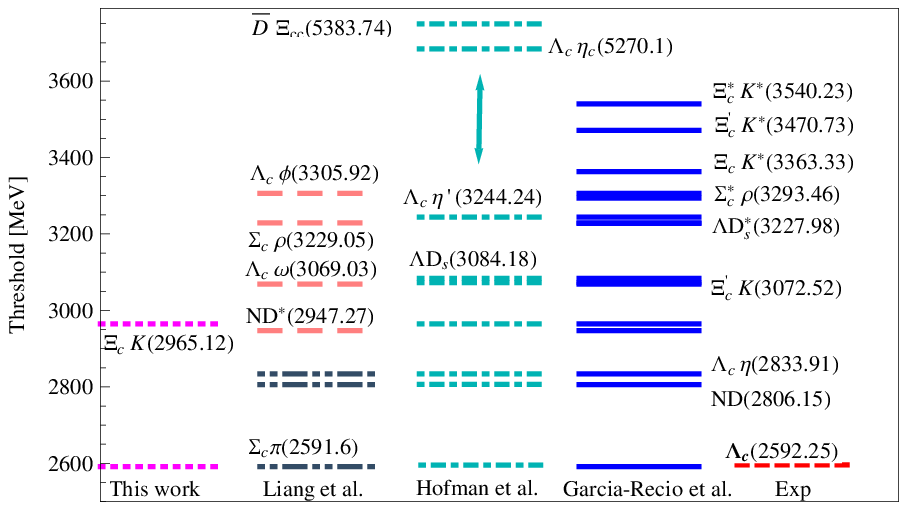}
\includegraphics[width=0.48\textwidth]{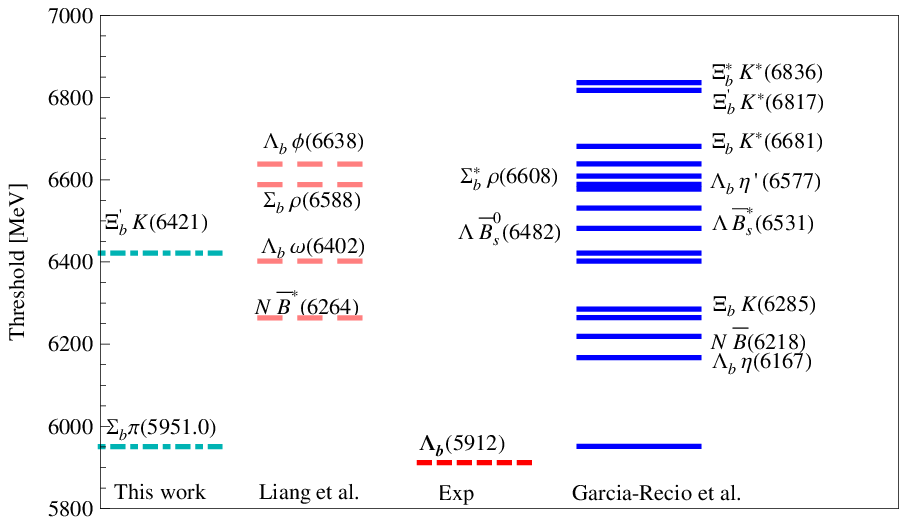}
\caption{Thresholds of the coupled channels considered in different works for the singly charmed and bottom baryon sector with $J^P=1/2^-$ and
$(S=0,I=0)$:  Liang \textit{et al.}~\cite{Liang:2014eba,Liang:2014kra}, Hofmann \textit{et al.}~\cite{Hofmann:2005sw},Garcia-Recio  \textit{et al.}~\cite{GarciaRecio:2008dp,GarciaRecio:2012db},and the experiment in Ref.~\cite{Agashe:2014kda}. In the
left figure, two model spaces denoted by dot-dot-dashed lines (pseudoscalar-baryon) and  dashed lines (vector-baryon), respectively, were  studied in Ref.~\cite{Liang:2014kra}. }
\label{Fig:threshold}
\end{figure}

As mentioned previously, compared to the studies of Refs.~\cite{GarciaRecio:2008dp,GarciaRecio:2012db,Liang:2014eba,Liang:2014kra}, we have only considered the minimum number of coupled channels dictated by chiral symmetry and its breaking.  Such an approach is only appropriate if close to the dynamically generated states no other coupled channels with the same quantum numbers exist. Otherwise, one may need to take into account those channels involving either vector mesons (light or heavy) or noncharmed (bottom) baryons. As can be seen from Fig.~\ref{Fig:threshold}, it is clear that for the dynamical generation of the $\Lambda_c(2595)$ and the $\Lambda_b(5912)$, our minimum coupled channel space indeed includes the most relevant channels, i.e., the $\Lambda_c \pi$ and
the $\Sigma_b \pi$, while the next-closest coupled channels excluded in our space,  the $N D$ and $N\bar{B}$, are roughly 200 MeV above.  On the other hand, the $\Lambda_c(2940)$ state is
close not only to the $\Xi'_c K$ channel taken into account in our framework but also to $\Lambda_c\omega$ and $\Lambda D_s$. As a result, our model space may be too restricted, and the result should be taken with care. This might be the reason why our prediction is about 100 MeV off the experimental mass of this resonance.

It has long been an important and challenging work to differentiate hadronic states of different nature, e.g.,
whether being composite states of other hadrons or being ``genuine'' (multi)quark states.
Half a century ago, Weinberg proposed the so-called compositeness condition, which allowed
him to tell that the deuteron is a weakly bound state of a proton and  a neutron, instead of a genuine six-quark state~\cite{Weinberg:1965zz}.
With renewed interests in hadron spectroscopy, there have been some recent works on this issue~\cite{Hanhart:2010wh,Baru:2003qq,Cleven:2011gp}.
Extensions to larger binding energies in the $s$ wave for bound states~\cite{Gamermann:2009uq} and resonances~\cite{YamagataSekihara:2010pj} and to
 higher partial waves for mesonic states \cite{Aceti:2012dd,Xiao:2012vv} and baryonic states~\cite{Aceti:2014ala,Aceti:2014wka} have been performed.\footnote{ See also Ref.~\cite{Sekihara:2014kya} and references cited therein.}

 Following Ref.~\cite{ Aceti:2014ala}, one can define the weight of a hadron-hadron component in a composite particle
 as
 \begin{equation}
 X_i=-\mathrm{Re}\left[g_i^2\left[\frac{\partial G_i^{II}(s)}{\partial \sqrt{s}}\right]_{\sqrt{s}=\sqrt{s_0}}\right]£¬
 \end{equation}
 where $\sqrt{s_0}$ is the pole position, $G_i^{II}$ is the loop function evaluated on the second Riemann sheet, and
 $g_i$ is the  couplings of the respective resonance or bound state to channel $i$ calculated  as
 \begin{equation}
 g_i^2=\mathop{\mathrm{lim}}_{\sqrt{s}\rightarrow\sqrt{s_0}}(\sqrt{s}-\sqrt{s_0})T_{ii}^{II},
 \end{equation}
 where $T_{ii}^{II}$ is the $ii$ element of the $T$ amplitude on the second Riemann sheet.

 The deviation of the sum of $X_i$ from unity is related to the energy dependence of the $s$-wave  potential,
 \begin{equation}
 \sum_i X_i=1-Z,
 \end{equation}
 where
 \begin{equation}
 Z=-\sum_{ij}\left[g_i G_i^{II}(\sqrt{s})\frac{\partial V_{ij}(\sqrt{s})}{\partial \sqrt{s}}G_j^{II}(\sqrt{s})g_j\right]_{\sqrt{s}=\sqrt{s}_0}.
 \end{equation}
 Although in certain cases $Z$ can be attributed to the weight of the missing channels (see, e.g., Ref.~\cite{Aceti:2014ala}),  it is not clear
how to interpret $Z$ obtained from the smooth energy dependence of the chiral potential $V$~\cite{Aceti:2014wka}. In addition, in a coupled-channel scenario, we noticed that
different treatments of the regularization schemes can reshuffle the contributions between $\sum_i X_i$ and $Z$, thus complicating the interpretation of the so-called compositeness~\cite{Geng:2015}.
To complicate things more,  for  processes involving short distances, it is the wave function at the origin that matters ($g_i G_i$ for the $s$ wave)~\cite{Gamermann:2009fv}. For an extensive discussion on
this issue, see  Ref.~\cite{Aceti:2014wka}, which concluded that to judge the relevance of each channel one has to study different physical processes. In the present context, we may similarly conclude that
the relevance of the channels neglected in the present work compared with those of Refs.~\cite{GarciaRecio:2008dp,GarciaRecio:2012db,Liang:2014eba,Liang:2014kra} can only be reliably evaluated in specific physical processes, which will be left for future studies.\footnote{In a recent work, the compositeness of the strange, charmed, and beauty $\Lambda$ states have been studied in
the extended Weinberg--Tomozawa framework supplemented by the SU(6) and the heavy-quark symmetries~\cite{Garcia-Recio:2015jsa}.}

\begin{table}[!htb]
      \centering
      \caption{$\phi B$ scattering lengths  $a$ (in units of fm) in the bottom sector with $J^P=1/2^-$.}\label{Table:scat3b}
      \begin{tabular}{clllclll}
         \hline\hline
          $(S,I)^{M}$ &     Channel   & $a$ &  $(S,I)^{M}$ &     Channel   & $a$ \\\hline
          $(1,\frac{1}{2})^{[\overline{3}]}$ & $\Lambda_{b}K$(6115.0)  & $-0.111$ &$(1,\frac{3}{2})^{[6]}$ & $\Sigma_{b}K$(6308.6) & $-0.138$  \\
          $(0,1)^{[\overline{3}]}$ & $\Xi_{b}K$(6285.1)  & $-0.239-i0.040$   & $(1,\frac{1}{2})^{[6]}$ & $\Sigma_{b}K$(6308.6) & $-0.419$  \\
          $(0,1)^{[\overline{3}]}$ & $\Lambda_{b}\pi$(5757.4)  & $0.003$   & $(0,2)^{[6]}$ & $\Sigma_{b}\pi$(5951.0) & $-0.102$  \\
          $(0,0)^{[\overline{3}]}$ & $\Xi_{b}K$(6285.1)  & $-0.204-i0.003$   & $(0,1)^{[6]}$ & $\Xi'_{b}K$(6421.6) & $-0.211-i0.007$  \\
          $(0,0)^{[\overline{3}]}$ & $\Lambda_{b}\eta$(6167.3)  & $-0.150$   & $(0,1)^{[6]}$ & $\Sigma_{b}\eta$(6360.9) & $-0.273-i0.014$  \\
          $(-1,\frac{3}{2})^{[\overline{3}]}$ & $\Xi_{b}\pi$(5927.5)  & $-0.067$   & $(0,1)^{[6]}$ & $\Sigma_{b}\pi$(5951.0) & $1.162$  \\
          $(-1,\frac{1}{2})^{[\overline{3}]}$ & $\Xi_{b}\pi$(5927.5)  & $-0.245$  & $(0,0)^{[6]}$ & $\Xi'_{b}K$(6421.6) & $-0.398-i0.019$  \\
          $(-1,\frac{1}{2})^{[\overline{3}]}$ & $\Xi_{b}\eta$(6337.4)  & $-0.208-i0.028$   & $(0,0)^{[6]}$ & $\Sigma_{b}\pi$(5951.0) & $-0.598$  \\
          $(-1,\frac{1}{2})^{[\overline{3}]}$ & $\Lambda_{b}\overline{K}$(6115.0)  & $-0.181-i0.206$  & $(-1,\frac{3}{2})^{[6]}$ & $\Xi'_{b}\pi$(6064.0) & $0.012$  \\
          $(-2,1)^{[\overline{3}]}$ & $\Xi_{b}\overline{K}$(6285.1)  & $-0.118$   & $(-1,\frac{3}{2})^{[6]}$ & $\Sigma_{b}\overline{K}$(6308.6) & $-0.350-i0.061$  \\
          $(-2,0)^{[\overline{3}]}$ & $\Xi_{b}\overline{K}$(6285.1)  & $-0.507$   & $(-1,\frac{1}{2})^{[6]}$ & $\Xi'_{b}\pi$(6064.0) & $0.497$  \\
          $(-1,\frac{1}{2})^{[6]}$ & $\Xi'_{b}\eta$(6473.9) & $-0.222-i0.020$    & $(-2,1)^{[6]}$ & $\Omega_{b}\pi$(6186.0) & $0.086$  \\
          $(-1,\frac{1}{2})^{[6]}$ & $\Omega_{b}K$(6543.6) & $-0.279-i0.014$    & $(-2,0)^{[6]}$ & $\Xi'_{b}\overline{K}$(6421.6) & $-0.214$  \\
          $(-1,\frac{1}{2})^{[6]}$ & $\Sigma_{b}\overline{K}$(6308.6) & $-0.185-i0.008$    & $(-2,0)^{[6]}$ & $\Omega_{b}\eta$(6595.9) & $-0.187-i0.003$  \\
          $(-2,1)^{[6]}$ & $\Xi'_{b}\overline{K}$(6421.6) & $-0.245-i0.112$    & $(-3,\frac{1}{2})^{[6]}$ & $\Omega_{b}\overline{K}$(6543.6) & $-0.153$  \\
         \hline\hline
      \end{tabular}
\end{table}

\begin{table}[!htb]
\small
      \centering
      \caption{$\phi B$ scattering lengths  $a$ (in units of fm) in the bottom sector with $J^P=3/2^-$.}\label{Table:scat3b2}
      \begin{tabular}{clllclll}
         \hline\hline
          $(S,I)^{M}$ &     Channel   & $a$ &  $(S,I)^{M}$ &     Channel   & $a$ \\\hline
          $(1,\frac{3}{2})^{[6^{*}]}$ & $\Sigma^{*}_{b}K$(6329.1)           & $-0.126$        & $(-1,\frac{1}{2})^{[6^{*}]}$  & $\Xi^{*}_{b}\pi$(6087.3)            &  $0.971$   \\
          $(1,\frac{1}{2})^{[6^{*}]}$ & $\Sigma^{*}_{b}K$(6329.1)           &  $-0.325$        & $(-1,\frac{1}{2})^{[6^{*}]}$  & $\Xi^{*}_{b}\eta$(6497.2)           & $-0.186-0.009i$  \\
          $(0,2)^{[6^{*}]}$           & $\Sigma^{*}_{b}\pi$(5971.5)         &  $-0.096$        & $(-1,\frac{1}{2})^{[6^{*}]}$  & $\Omega^{*}_{b}K$(6564.6)           &  $-0.233-0.008i$   \\
          $(0,1)^{[6^{*}]}$           & $\Xi^{*}_{b}K$(6444.9)              &  $-0.183-0.004i$ & $(-1,\frac{1}{2})^{[6^{*}]}$  & $\Sigma^{*}_{b}\overline{K}$(6329.1)& $-0.151-0.024i$   \\
          $(0,1)^{[6^{*}]}$           & $\Sigma^{*}_{b}\eta$(6381.3)        & $-0.223-0.007i$ & $(-2,1)^{[6^{*}]}$            & $\Xi^{*}_{b}\overline{K}$(6444.9)   &  $-0.222-0.056i$   \\
          $(0,1)^{[6^{*}]}$           & $\Sigma^{*}_{b}\pi$(5971.5)         &  $8.412$         & $(-2,1)^{[6^{*}]}$            & $\Omega^{*}_{b}\pi$(6207.0)         &  $0.109$  \\
          $(0,0)^{[6^{*}]}$           & $\Xi^{*}_{b}K$(6444.9)              &  $-0.312-0.009i$ & $(-2,0)^{[6^{*}]}$            & $\Xi^{*}_{b}\overline{K}$(6444.9)   &  $-0.184$   \\
          $(0,0)^{[6^{*}]}$           & $\Sigma^{*}_{b}\pi$(5971.5)         &  $-0.425$        & $(-2,0)^{[6^{*}]}$            & $\Omega^{*}_{b}\eta$(6616.9)        &  $-0.165-0.002i$   \\
          $(-1,\frac{3}{2})^{[6^{*}]}$& $\Xi^{*}_{b}\pi$(6087.3)            &  $0.042$         & $(-3,\frac{1}{2})^{[6^{*}]}$  & $\Omega^{*}_{b}\overline{K}$(6564.6)&  $-0.139$   \\
          $(-1,\frac{3}{2})^{[6^{*}]}$& $\Sigma^{*}_{b}\overline{K}$(6329.1)& $-0.276-0.029i$ &                           &                  &    \\
         \hline\hline
      \end{tabular}
\end{table}

\begin{table}[!htb]
\scriptsize
      \centering
      \caption{$\phi B$ scattering lengths $a$ (in units of fm) in the charmed sector with $J^P=1/2^-$.}\label{Table:scat3c}
      \begin{tabular}{clllclll}
         \hline\hline
          $(S,I)^{M}$ &     Channel   & \multicolumn{2}{c}{$a$ }                     & $(S,I)^{M}$ &     Channel   & \multicolumn{2}{c}{$a$}  \\
                   &                                  &  Present work & Ref.~\cite{Liu:2012uw} &                     &                       & Present work &Ref.~\cite{Liu:2012uw}  \\\hline
          $(1,\frac{1}{2})^{[\overline{3}]}$ & $\Lambda_{c}K$  & $-0.135$ & $-0.032\pm0.038$ & $(1,\frac{3}{2})^{[6]}$ & $\Sigma_{c}K$ & $-0.181$& $-0.44$$\mp$0.23 \\
          $(0,1)^{[\overline{3}]}$ & $\Xi_{c}K$  & $-0.281-i0.308$ & $0.77+i0.18$ &  $(1,\frac{1}{2})^{[6]}$ & $\Sigma_{c}K$ & $-8.114$& 0.62$\pm$0.12 \\
          $(0,1)^{[\overline{3}]}$ & $\Lambda_{c}\pi$  & 0.002 & 0.006 &  $(0,2)^{[6]}$ & $\Sigma_{c}\pi$ & $-0.119$& $-0.25$$\mp$0.031 \\
          $(0,0)^{[\overline{3}]}$ & $\Xi_{c}K$  & $-0.338-i0.020$ & 0.99$\pm$0.076&  $(0,1)^{[6]}$ & $\Xi'_{c}K$ & $-0.365-i0.097$& $(0.18+i0.37)\mp$0.12 \\
          $(0,0)^{[\overline{3}]}$ & $\Lambda_{c}\eta$  & $-0.281$ & $(0.35+i0.19)\pm0.044$&  $(0,1)^{[6]}$ & $\Sigma_{c}\eta$ & $-0.787-i0.942$& $(0.18+i0.2)\mp$0.034 \\
          $(-1,\frac{3}{2})^{[\overline{3}]}$ & $\Xi_{c}\pi$  & $-0.072$ & $-0.11$$\pm$0.0052 & $(0,1)^{[6]}$ & $\Sigma_{c}\pi$ & 0.376& 0.28 \\
          $(-1,\frac{1}{2})^{[\overline{3}]}$ & $\Xi_{c}\pi$  & 1.600 & 0.32$\pm$0.0052 & $(0,0)^{[6]}$ & $\Xi'_{c}K$ & $-1.361-i1.174$& $(1.4+i0.56)\mp$0.12 \\
          $(-1,\frac{1}{2})^{[\overline{3}]}$ & $\Xi_{c}\eta$  & $-0.266-i0.197$ & $(0.54+i0.098)\pm0.011$ & $(0,0)^{[6]}$ & $\Sigma_{c}\pi$ & $-28.204$& 0.65$\mp$0.078 \\
          $(-1,\frac{1}{2})^{[\overline{3}]}$ & $\Lambda_{c}\overline{K}$  & $-0.237-i0.148$ & $(0.79+i0.27)\pm0.038$ & $(-1,\frac{3}{2})^{[6]}$ & $\Xi'_{c}\pi$ & $-0.025$& $-0.19$$\mp$0.016 \\
          $(-2,1)^{[\overline{3}]}$ & $\Xi_{c}\overline{K}$  & $-0.141$ & $-0.028$$\pm$0.038 & $(-1,\frac{3}{2})^{[6]}$ & $\Sigma_{c}\overline{K}$ & $0.141-i0.770$& $0.12+i0.37$ \\
          $(-2,0)^{[\overline{3}]}$ & $\Xi_{c}\overline{K}$  & 12.014 & 1.7$\mp$0.038 & $(-1,\frac{1}{2})^{[6]}$ & $\Xi'_{c}\pi$ & $-0.022$& 0.23$\mp$0.016 \\
          $(-1,\frac{1}{2})^{[6]}$ & $\Xi'_{c}\eta$ & $-0.196-i0.269$ & $(0.55+i0.49)\mp0.24$ & $(-2,1)^{[6]}$ & $\Omega_{c}\pi$ & 0.046& $-0.062$ \\
          $(-1,\frac{1}{2})^{[6]}$ & $\Omega_{c}K$ & $-0.508-i0.160$  & $(1.4+i0.56)\mp0.23$ & $(-2,0)^{[6]}$ & $\Xi'_{c}\overline{K}$ & $-0.413$& 0.61$\mp$0.12 \\
          $(-1,\frac{1}{2})^{[6]}$ & $\Sigma_{c}\overline{K}$ & $-0.345-i0.013$ & $(2.0+i0.092)\mp0.35$ & $(-2,0)^{[6]}$ & $\Omega_{c}\eta$ & $-0.277-i0.015$& $(0.68+i0.4)\mp$0.14 \\
          $(-2,1)^{[6]}$ & $\Xi'_{c}\overline{K}$ & $-0.088-i0.168$ & $(-0.11+i0.37)\mp 0.12$ & $(-3,\frac{1}{2})^{[6]}$ & $\Omega_{c}\overline{K}$ & $-0.197$& $-0.33$$\mp$0.23 \\
         \hline\hline
      \end{tabular}
\end{table}

\begin{table}[!htb]
\scriptsize
      \centering
      \caption{$\phi B$ scattering lengths  $a$ (in units of fm) in the charmed sector with $J^P=3/2^-$.}\label{Table:scat3c2}
      \begin{tabular}{clllclll}
         \hline\hline
          $(S,I)^{M}$ &     Channel   & \multicolumn{2}{c}{$a$ }                     & $(S,I)^{M}$ &     Channel   & \multicolumn{2}{c}{$a$}  \\
                   &                                  &  Present work & Ref.~\cite{Liu:2012uw} &                     &                       & Present work &Ref.~\cite{Liu:2012uw}  \\\hline
          $(1,\frac{3}{2})^{[6^{*}]}$ & $\Sigma^{*}_{c}K$(3013.5)           & $-0.147$       &$-0.45\mp0.23$              & $(-1,\frac{1}{2})^{[6^{*}]}$  & $\Xi^{*}_{c}\pi$(2783.9)            & $ 0.459$       &$(0.23-0.027i)\mp0.016$  \\
          $(1,\frac{1}{2})^{[6^{*}]}$ & $\Sigma^{*}_{c}K$(3013.5)           & $-0.683$        &$0.63\mp0.12$               & $(-1,\frac{1}{2})^{[6^{*}]}$  & $\Xi^{*}_{c}\eta$(3193.8)           & $-0.263-0.060i$&$(0.57+0.5i)\mp0.24$  \\
          $(0,2)^{[6^{*}]}$           & $\Sigma^{*}_{c}\pi$(2655.9)         & $-0.104$       &$-0.25\mp0.031$             & $(-1,\frac{1}{2})^{[6^{*}]}$  & $\Omega^{*}_{c}K$(3261.5)           & $-0.324-0.036i$&$(1.4+0.56i)\mp0.23$   \\
          $(0,1)^{[6^{*}]}$           & $\Xi^{*}_{c}K$(3141.5)              & $-0.246-0.020i$&$(0.13+0.37i)\mp0.12$       & $(-1,\frac{1}{2})^{[6^{*}]}$  & $\Sigma^{*}_{c}\overline{K}$(3013.5)& $-0.222-0.009i$&$(2.0+0.092i)\mp0.35$   \\
          $(0,1)^{[6^{*}]}$           & $\Sigma^{*}_{c}\eta$(3065.8)        & $-0.398-0.059i$&$(0.16+0.2i)\mp0.034$       & $(-2,1)^{[6^{*}]}$            & $\Xi^{*}_{c}\overline{K}$(3141.5)   & $-0.199-0.211i$&$(-0.12+0.37i)\mp0.12$   \\
          $(0,1)^{[6^{*}]}$           & $\Sigma^{*}_{c}\pi$(2655.9)         & $0.820$        &$0.27-0.021i$               & $(-2,1)^{[6^{*}]}$            & $\Omega^{*}_{c}\pi$(2903.9)         & $0.085$        &$-0.062$    \\
          $(0,0)^{[6^{*}]}$           & $\Xi^{*}_{c}K$(3141.5)              & $-0.572-0.072i$&$(1.5+0.56i)\mp0.12$        & $(-2,0)^{[6^{*}]}$            & $\Xi^{*}_{c}\overline{K}$(3141.5)   & $-0.243$       &$(0.52-0.0032i)\mp0.12$\\
          $(0,0)^{[6^{*}]}$           & $\Sigma^{*}_{c}\pi$(2655.9)         & $-0.761$       &$(0.67+0.032i)\mp0.078$     & $(-2,0)^{[6^{*}]}$            & $\Omega^{*}_{c}\eta$(3313.8)        & $-0.201-0.005i$&$(0.64+0.4i)\mp0.14$   \\
          $(-1,\frac{3}{2})^{[6^{*}]}$& $\Xi^{*}_{c}\pi$(2783.9)            & $ 0.022$       &$(-0.19-0.0027i)\mp0.016$   & $(-3,\frac{1}{2})^{[6^{*}]}$  & $\Omega^{*}_{c}\overline{K}$(3261.5)& $-0.157$       &$-0.34\mp0.23$\\
          $(-1,\frac{3}{2})^{[6^{*}]}$& $\Sigma^{*}_{c}\overline{K}$(3013.5)& $-0.539-0.242i$&$0.13+0.37i$                &                           &                  &   &  \\
         \hline\hline
      \end{tabular}
\end{table}

\subsection{Scattering lengths}
Scattering lengths provide vital information on the strong interactions. Although direct experimental measurements of the scattering lengths between
a charmed (bottom) baryon and a pseudoscalar meson cannot be foreseen in the near future, rapid developments in lattice QCD  may soon fill the gap.
In Tables~\ref{Table:scat3b}, \ref{Table:scat3b2}, \ref{Table:scat3c}, and \ref{Table:scat3c2}, we tabulate the scattering lengths calculated in the dimensional regularization scheme, defined as
\begin{equation}
a_{jj}=-\frac{M_j}{4\pi(M_j+m_j)}T^{(S,I)}_{jj},
\end{equation}
for channel $j$ with strangeness $S$ and isospin $I$, where $M_j$ and $m_j$ are the respective baryon and meson masses of that channel. For the sake of comparison, we list the chiral perturbation theory results
of Ref.~\cite{Liu:2012uw}. One should note, however, that Ref.~\cite{Liu:2012uw} calculated the scattering lengths up to $\mathcal{O}(p^3)$ .
While in our study, only the leading-order ($\mathcal{O}(p)$) chiral perturbation theory kernel is used, and in addition we work with the UChPT.

Examining the scattering lengths in the charmed sector, we notice that because of the existence of a bound state just below their respective thresholds, the scattering lengths
for the $\Sigma_c K$ channel with $(S,I)^M=(1,1/2)^{[6]}$ and for the $\Sigma_c\pi$ channel with $(S,I)^M=(0,0)^{[6]}$ are quite large and negative, i.e., $a_{\Sigma_c K}=-8.114$ and $a_{\Sigma_c\pi}=-28.204$.
Therefore, a future lattice QCD study of these two channels may be able to test to what extent  the scenario of these states being dynamically generated is true.

\section{Exploratory NLO study of the $1/2^-$ sector}

In this section, we study the effects of the NLO potentials. In principle, higher-order effects in the UChPT can be taken into account systematically if relevant low-energy constants (LECs) can be fixed reliably.  However, this is not the case in the present study. Therefore, we will turn to some phenomenological means to fix
some of the LECs and vary others within their natural range to study the effects of the NLO potentials. As an exploratory study, we limit ourselves to the $1/2^-$ sector.

To reduce the number of unknown LECs, we use the following NLO Lagrangians in the heavy-meson formulation~\cite{Liu:2012uw}:
\begin{eqnarray}\label{Eq:LOLag2}
        \begin{split}
           \mathcal{L}_{H\Phi}^{(2)}
           & = \bar{c}_{0}Tr[\bar{H}_{\bar{3}}H_{\bar{3}}]Tr[\chi_{+}] + \bar{c}_{1}Tr[\bar{H}_{\bar{3}}\tilde{\chi}_{+}H_{\bar{3}}] + (\bar{c}_{2}-\frac{2g_{6}^{2}+g_{2}^{2}}{4M_{0}})Tr[\bar{H}_{\bar{3}}v \cdot uv \cdot uH_{\bar{3}}]\\
           & +(\bar{c}_{3}-\frac{2g_{6}^{2}-g_{2}^{2}}{4M_{0}})\bar{H}_{\bar{3}}^{ab}v \cdot u_{a}^{c}v \cdot u_{b}^{d}H_{\bar{3},cd} + c_{0}Tr[\bar{H}_{6}H_{6}]Tr[\chi_{+}] + c_{1}Tr[\bar{H}_{6}\tilde{\chi}_{+}H_{6}]\\
           & +(c_{2}-\frac{2g_{2}^{2}+g_{1}^{2}}{4M_{0}})Tr[\bar{H}_{6}v \cdot uv \cdot uH_{6}] + (c_{3}+\frac{2g_{2}^{2}-g_{1}^{2}}{4M_{0}})\bar{H}_{6}^{ab}v \cdot u_{a}^{c}v \cdot u_{b}^{d}H_{6,cd}\\
           & +c_{4}Tr[\bar{H}_{6}H_{6}]Tr[v \cdot uv \cdot u] ,
        \end{split}
\end{eqnarray}
where $\chi_{+}$ and $\tilde{\chi}_{+}$ are defined as 
\begin{eqnarray}\label{chi}
      \begin{split}
          \chi_{\pm} & = \xi^{+}\chi\xi^{+} \pm \xi\chi\xi \\
          \chi & = \mathrm{diag}(m_{\pi}^{2},m_{\pi}^{2},2m_{K}^{2}-m_{\pi}^{2})\\
          \tilde{\chi}_{\pm} & = \chi_{\pm}-\frac{1}{3}Tr[\chi_{\pm}],
      \end{split}
\end{eqnarray}
with $\xi=\exp(i\frac{\phi}{2f})$.

The LECs $g_2$ and $g_4$ can be fixed by reproducing the $\Sigma_c$ and $\Sigma_c^*$ widths, while the other $g_{i}$'s can be
related to them using either quark model symmetries or the heavy-quark spin symmetry.  The LECs $\bar{c}_{i}$ and $c_{i}$ are fixed using the (broken) SU(4) symmetry in Ref.~\cite{Liu:2012uw}. In the present work, we follow Ref.~\cite{Liu:2012uw} and use the values determined there and reproduced in Tables \ref{Table:constantg}, \ref{Table:constant1},  and \ref{Table:constant2}.~\footnote{The values are
slightly different from those of Ref.~\cite{Liu:2012uw} because some relations among the LECs are stated incorrectly there.} The LEC $\alpha'$ cannot be determined, and we will estimate its contribution below assuming a natural value within the range of $-1\sim 1$ as in Ref.~\cite{Liu:2012uw}.

\begin{table}[t]
      \renewcommand{\arraystretch}{1.6}
      \setlength{\tabcolsep}{0.55cm}
      \centering
      \caption{Constants in Eq.~(\ref{Eq:LOLag2}) for the antitriplet.}\label{Table:constantg}
      \begin{tabular}{cccccc}
         \hline\hline
          $|g_{1}|$ & $|g_{2}|$ & $|g_{3}|$ & $|g_{4}|$ & $|g_{5}|$ & $|g_{6}|$\\
         \hline
          0.98 & 0.60 & 0.85 & 1.0 & 1.5 & 0 \\
         \hline\hline
      \end{tabular}
\end{table}
\begin{table}[h]
      \renewcommand{\arraystretch}{1.6}
      \setlength{\tabcolsep}{0.55cm}
      \centering
      \caption{Constants in Eq.~(\ref{Eq:LOLag2}) for the antitriplet (in units of $GeV^{-1}$).}\label{Table:constant1}
      \begin{tabular}{llll}
         \hline\hline
          $\bar{c}_{0}$ & $\bar{c}_{1}$ & $\bar{c}_{2}$ & $\bar{c}_{3}$\\\hline
          $-0.32$ & $-0.52$ & $-1.78+\frac{1}{3}\frac{\alpha'}{4\pi f}$ & $-0.03-\frac{1}{3}\frac{\alpha'}{4\pi f}$\\
         \hline\hline
      \end{tabular}
\end{table}
\begin{table}[h]
      \renewcommand{\arraystretch}{1.6}
      \setlength{\tabcolsep}{0.55cm}
      \centering
      \caption{Constants in Eq.~(\ref{Eq:LOLag2}) for the sextet (in units of $GeV^{-1}$).}\label{Table:constant2}
      \begin{tabular}{lllll}
         \hline\hline
          $c_{0}$ & $c_{1}$ & $c_{2}$ & $c_{3}$ & $c_{4}$\\
         \hline
          $-0.61$ & $-0.98$ & $-2.07-2\frac{\alpha'}{4\pi f}$ & $-0.84$ & $\frac{\alpha'}{4\pi f}$\\
         \hline\hline
      \end{tabular}
\end{table}

In the NLO study, we fix the subtraction constant in the same way as in the LO case and search for poles on the complex plane. The results  are tabulated in
Tables \ref{Table:csub1} and \ref{Table:bsub1}.

Compared to the LO case, we find some substantial changes when the NLO potentials are taken into account. For instance, in the charmed sector,
one dynamically generated state in the antitriplet sector disappears while a new one appears with $\alpha'=-1.0$.  In the bottom sector,
one dynamically generated state in the antitriplet sector disappears as well. This implies that the NLO chiral potential  has a huge impact on the  predicted states  in the antitriplet sector.

 In the sextet sector, on the other hand, the changes are more moderate. Most states move a few tens of MeV
compared to their LO counterparts with a few exceptions.  However, the unknown LEC $\alpha'$ seems to affect the predictions a lot. In particular, when $\alpha'=-1$, many states disappear. Clearly, from the comparison with the LO results, we may conclude that $\alpha'=-1$ is not preferred.

One of the possible reasons why the results in the antitriplet sector change more dramatically than those in the sextet sector is the following. In the sextet sector, we have refitted the subtraction constant to produce the states at $(2591,0)$ and $(5912,0)$ MeV, while no such readjustments have been made for the antitriplet sectors. Nevertheless, one should note that even at NLO the $\Lambda_c(2595)$ and $\Lambda_b(5912)$ appears naturally as
dynamically generated states without the need for an unnatural subtraction constant.

In fact, due to the lack of enough experimental information to have good control on the NLO LECs, none of the above observations is surprising. In Ref.~\cite{Liu:2012uw}, Liu and Zhu already found that in many cases the NLO potentials are larger than the LO ones (see Tables I, II, and III of their paper).  Our studies confirmed their findings and showed that some of the LO predictions are subject to substantial modifications while some others may remain relatively stable. More experimental or  lattice QCD inputs are clearly needed to check the results and clarify the situation. On one hand, one needs to be cautious about those results where higher-order potentials are shown to be particularly relevant. On the other hand, one should note that the NLO contributions depend critically on the way the relevant
LECs are estimated. If we had put them equal to zero, the contributions would vanish. Clearly, the LECs should be determined in a more reliable way in order to study the effects of higher-order potentials.

\begin{table}[!htb]
      \centering
      \caption{Dynamically generated charmed baryons of $J^P=1/2^-$  in the NLO UChPT in comparison with those in the LO.
At NLO, three values for the LEC $\alpha'$ have been taken. The subtraction constant is fixed in the same way as in the LO case.  All energies are in units of MeV and $(S,I)^M$ denotes (strangeness, isospin)$^\mathrm{SU(3) multiplet}$.}\label{Table:csub1}
      \begin{tabular}{lllllll}
         \hline\hline
         \multirow{2}*{LO} & \multicolumn{3}{c}{NLO}  & \multirow{2}*{$(S,I)^{M}$} & Main channels  & \multirow{2}*{Ref.~\cite{Agashe:2014kda}} \\
                           & $\alpha'=0$ & $\alpha'=1.0$ & $\alpha'=-1.0$ & & (threshold) & \\\hline
         -          & - & - & $(2936,-i15)$ & $(0,1)^{[\overline{3}]}$ & $\Xi_{c}K$(2965.1)  \\
         $(2721,0)$ & $(2807,0)$ & $(2794,0)$ & $(2820,0)$ & $(0,0)^{[\overline{3}]}$ & $\Xi_{c}K$(2965.1)  \\
         $(2623,-i12)$ & -& -& -&  $(-1,\frac{1}{2})^{[\overline{3}]}$ & $\Lambda_{c}\overline{K}$(2782.1),$\Xi_{c}\pi$(2607.5)  \\
         $(2965,0)$ & $(2736,0)$ & $(2741,0)$ & $(2732,0)$ &  $(-2,0)^{[\overline{3}]}$ & $\Xi_{c}\overline{K}$(2965.1)  \\\hline

         $(2948,0)$ & $(2918,0)$ & $(2848,0)$ & - &  $(1,\frac{1}{2})^{[6]}$ & $\Sigma_{c}K$(2949.1)  \\
         $(2674,-i51)$ & $(2699,-i107)$ & $(2702,-i102)$ & $(2699,-i105)$ & $(0,1)^{[6]}$ & $\Sigma_{c}\pi$(2591.5)  \\
         $(2999,-i16)$ & $(2985,-i0.01)$ & $(2984,-i33)$ & -& $(0,1)^{[6]}$ & $\Sigma_{c}\eta$(3001.4),$\Xi'_{c}\overline{K}$(3072.4)  &  \\
         $(2591,0)$ & $(2591,0)$ & $(2591,0)$ & $(2591,0)$ & $(0,0)^{[6]}$ & $\Sigma_{c}\pi$(2591.5) & $\Lambda_c(2595)$  \\
         $(3069,-i12)$ & $(3025,-i19)$ & $(2954,-i11)$ &- & $(0,0)^{[6]}$ & $\Xi'_{c}K$(3072.4)   & $\Lambda_c(2940)?$\\
         $\underline{(2947,-i34)}$ & $(2925,-i13)$ & $(2857,-i11)$ & -& $( -1,\frac{3}{2})^{[6]}$ & $\Sigma_{c}\overline{K}$(2949.1)  \\
         $(2695,0)$ & $(2567,0)$ & $(2568,0)$ & $(2565,0)$ & $(-1,\frac{1}{2})^{[6]}$ & $\Sigma_{c}\overline{K}$(2949.1) \\
         $(2827,-i55)$ & $(2824,-i74)$ & $(2813,-i70)$ & $(2836,-i75)$ & $(-1,\frac{1}{2})^{[6]}$ & $\Xi'_{c}\pi$(2714.7)  & $\Xi_c(2790)$?\\
         $\underline{(3123,-i44)}$ & $(3084,-i26)$ & $(3038,-i8)$ & -& $(-1,\frac{1}{2})^{[6]}$ & $\Omega_{c}K$(3190.8) \\
         -&  -& $(3005,-i38)$ &- & $(-2,1)^{[6]}$ & $\Xi'_{c}\overline{K}$(3072.4)  \\
         $(2946,0)$ & $(2815,0)$ & $(2821,0)$ & $(2809,0)$ & $(-2,0)^{[6]}$ & $\Xi'_{c}\overline{K}$(3072.4),$\Omega_{c}\eta$(3243.1)  \\
         \hline\hline
      \end{tabular}
\end{table}

\begin{table}[!htb]
      \centering
      \caption{The same as Table \ref{Table:csub1} but for the bottom baryons.}\label{Table:bsub1}
      \begin{tabular}{lllllll}
         \hline\hline
         \multirow{2}*{LO} & \multicolumn{3}{c}{NLO}  & \multirow{2}*{$(S,I)^{M}$} & Main channels  & \multirow{2}*{Ref.~\cite{Agashe:2014kda}} \\
                           & alpha=0 & alpha=1.0 & alpha=-1.0 & & (threshold) & \\\hline
         $(6082,-i57)$ & $(5888,-i13)$ & $(5890,-i13)$&$(5886,-i13)$ &  $(0,1)^{[\overline{3}]}$ & $\Xi_{b}K$(6285.1)  \\
         $(5922,0)$ & $(6042,0)$ &$(6032,0)$ & $(6050,0)$ & $(0,0)^{[\overline{3}]}$ & $\Xi_{b}K$(6285.1)  \\
         $(5869,0)$ & -& -& -& $(-1,\frac{1}{2})^{[\overline{3}]}$ & $\Lambda_{b}\overline{K}$(6115.0)  \\
         $(6118,-i51)$ & $(6026,-i61)$ & $(6023,-i56)$ & $(6028,-i68)$ & $(-1,\frac{1}{2})^{[\overline{3}]}$ & $\Xi_{b}\eta$(6337.4) \\
         $(6202,0)$ & $(5848,0)$& $(5848,0)$& $(5848,0)$& $(-2,0)^{[\overline{3}]}$ & $\Xi_{b}\overline{K}$(6285.1)  \\\hline

         $(6201,0)$ & $(6193,0)$ & $(6125,0)$ &- & $(1,\frac{1}{2})^{[6]}$ & $\Sigma_{b}K$(6308.6)  \\
         $(5967,-i9)$ & $(5928,0)$ & $(5926,0)$ & $(5930,0)$ & $(0,1)^{[6]}$ & $\Xi'_{b}K$(6421.6),$\Sigma_{b}\pi$(5951.0)   \\
         $(6223,-i14)$ & $(6215,-i3)$ & $(6160,-i1)$ & -& $(0,1)^{[6]}$ & $\Sigma_{b}\eta$(6360.9),$\Xi'_{b}K$(6421.6)  \\
         $(5912,0)$ & $(5912,0)$ & $(5912,0)$ & $(5912,0)$ & $(0,0)^{[6]}$ & $\Sigma_{b}\pi$(5951.0) $\Xi'_{b}K$(6421.6) & $\Lambda_b(5912)$\\
         $(6307,-i12)$ & $(6298,-i16)$ & $(6228,-i10)$ &- & $(0,0)^{[6]}$ & $\Xi'_{b}K$(6421.6) \\
         $(6213,-i25)$ & $(6189,-i6)$ & $(6123,-i2)$ &  -& $( -1,\frac{3}{2})^{[6]}$ & $\Sigma_{b}\overline{K}$(6308.6)  \\
         $(5955,0)$ & -& -&- & $( -1,\frac{1}{2})^{[6]}$ & $\Sigma_{b}\overline{K}$(6308.6)  \\
         $(6101,-i15)$ & $(6089,-i15)$ & $(6084,-i13)$ & $(6094,-i16)$ & $( -1,\frac{1}{2})^{[6]}$ & $\Xi'_{b}\pi$(6064.0),$\Omega_{b}K$(6543.6)  \\
         $(6364,-i27)$ & $(6345,-i18)$ & $(6295,-i5)$& -& $(-1,\frac{1}{2})^{[6]}$ & $\Omega_{b}K$(6543.6) \\
         $(6361,-i59)$ & $(6340,-i28)$ & $(6278,-i14)$ & -& $(-2,1)^{[6]}$ & $\Omega_{b}\pi$(6186.0),$\Xi'_{b}\overline{K}$(6421.6)  \\
         $(6169,0)$ & $(5984,0)$&$(5985,0)$ &$(5984,0)$ &  $(-2,0)^{[6]}$ & $\Xi'_{b}\overline{K}$(6421.6) \\
         \hline\hline
      \end{tabular}
\end{table}

\section{Summary and outlook}
We have studied the interaction between a singly charmed (bottom) baryon and a pseudoscalar meson in the unitarized chiral perturbation theory
using leading-order chiral Lagrangians. It is shown that the interactions are strong enough to generate a number of dynamically generated states. Some of them
can be naturally assigned to their experimental counterparts, such as the $\Lambda_b(5912)$ [$\Lambda_b^*(5920)$] and the $\Lambda_c(2595)$ [$\Lambda_c^*(2625)$]. By slightly fine-tuning the subtraction constant
in the dimensional regularization scheme or the cutoff value in the cutoff regularization scheme so that the masses of these states are produced,
we predict a number of additional states, the experimental counterparts of which remain unknown. We strongly encourage future experiments to search for these states.

In anticipation of future lattice QCD simulations of scattering lengths, as already happened in the light-baryon sector or the heavy-meson sector, we have calculated the scattering lengths
between the charmed (bottom) baryons and the pseudoscalar mesons.  A comparison between our results and those of the $\mathcal{O}(p^3)$ chiral perturbation theory study confirmed that there
is indeed strong attraction in some of the coupled channels, which hints at the possible existence of dynamically generated states.

 In future, the effects of higher-order potentials in the unitarized chiral perturbation theory need to be studied more carefully once relevant experimental or lattice QCD data become available. It should be noted, however, that
 the $\Lambda_c(2595)$ and $\Lambda_b(5912)$ and their $J^P=3/2^-$ counterparts seem to qualify as dynamically generated states even at next-to-leading order in the unitarized chiral perturbation theory.

\section{Acknowledgements}

 This work is supported in part  by the National Natural Science Foundation of China under Grants No. 11375024, No. 11205011, and No. 11105126;
 the New Century Excellent Talents in University Program of Ministry of Education of China under
Grant No. NCET-10-0029; and the Fundamental Research Funds for the Central Universities.

\section{Appendix: Clebsch--Gordan coefficients}
In this section, we tabulate the Clebsch--Gordan coefficients appearing in Eq.~(\ref{Eq:LOKernel}) for the
antitriplet (Tables \ref{Table:coeff3f} to \ref{Table:coeff3l}) and sextet (Tables \ref{Table:coeff6f} to \ref{Table:coeff6l}) ground-state charmed or bottom baryons interacting with the pseudoscalar mesons.

\begin{table}[!htp]
   \caption{$(S=1,I=1/2)$\label{Table:coeff3f}}

      \begin{tabular}{c|c}
         \hline\hline
           & $\Lambda_{c}$K \\
         \hline
         $\Lambda_{c}$K\ & 1\\
         \hline\hline
      \end{tabular}
\end{table}

\begin{table}[!hbp]
   \caption{$(S=0,I=1)$}
      \begin{tabular}{c|cc}
         \hline\hline
           & $\Xi_{c}$K & $\Lambda_{c}\pi$\\
         \hline
         $\Xi_{c}$K & 0 & $1$\\
         $\Lambda_{c}\pi$ & $1$ & 0 \\
         \hline\hline
      \end{tabular}
  \end{table}

\begin{table}[!hbp]
   \caption{$(S=0,I=0)$}
      \begin{tabular}{c|cc}
         \hline\hline
           & $\Xi_{c}$K & $\Lambda_{c}\eta$\\
         \hline
         $\Xi_{c}$K & $-2$ & $-\sqrt{3}$\\
         $\Lambda_{c}\eta$ & $-\sqrt{3}$ & 0 \\
         \hline\hline
      \end{tabular}
  \end{table}

\begin{table}[!hbp]
\caption{$(S=-1,I=3/2)$}
      \begin{tabular}{c|c}
         \hline\hline
           & $\Xi_{c}\pi$ \\
         \hline
         $\Xi_{c}\pi$ & 1 \\
         \hline\hline
      \end{tabular}
   \end{table}

\begin{table}[!hbp]
  \caption{$(S=-1,I=1/2)$}
      \begin{tabular}{c|ccc}
         \hline\hline
           & $\Xi_{c}\pi$ & $\Xi_{c}\eta$ & $\Lambda_{c}\overline{K}$\\
         \hline
         $\Xi_{c}\pi$ & $-2$ & 0 & $\sqrt{3/2}$ \\
         $\Xi_{c}\eta$ & 0 & 0 & $-\sqrt{3/2}$ \\
         $\Lambda_{c}\overline{K}$ & $\sqrt{3/2}$ & $-\sqrt{3/2}$ & $-1$ \\
         \hline\hline
      \end{tabular}
   \end{table}

\begin{table}[!hbp]
   \caption{$(S=-2,I=1)$}
         \begin{tabular}{c|c}
         \hline\hline
           & $\Xi_{c}\overline{K}$ \\
         \hline
         $\Xi_{c}\overline{K}$ & 1\\
         \hline\hline
      \end{tabular}
\end{table}

\begin{table}[!htb]
      \caption{$(S=-2,I=0)$\label{Table:coeff3l}}
      \begin{tabular}{c|c}
         \hline\hline
           & $\Xi_{c}\overline{K}$ \\
         \hline
         $\Xi_{c}\overline{K}$ & $-1$\\
         \hline\hline
      \end{tabular}
\end{table}


\begin{table}[!htb]
\caption{$(S=1,I=3/2)$\label{Table:coeff6f}}
      \begin{tabular}{c|c}
         \hline\hline
           & $\Sigma_{c}$K \\
         \hline
         $\Sigma_{c}$K & $2$\\
         \hline\hline
      \end{tabular}
 \end{table}

\begin{table}[!hbp]
      \caption{$(S=1,I=1/2)$}
         \begin{tabular}{c|c}
         \hline\hline
           & $\Sigma_{c}$K \\
         \hline
         $\Sigma_{c}$K & $-1$\\
         \hline\hline
      \end{tabular}
  \end{table}

\begin{table}[!hbp]
      \caption{$(S=0,I=2)$}
      \begin{tabular}{c|c}
         \hline\hline
           & $\Sigma_{c}\pi$ \\
         \hline
         $\Sigma_{c}\pi$ & $2$\\
         \hline\hline
      \end{tabular}
  \end{table}

\begin{table}[!hbp]
      \caption{$(S=0,I=1)$}
         \begin{tabular}{c|ccc}
         \hline\hline
           & $\Xi^{'}_{c}K$ & $\Sigma_{c}\eta$ & $\Sigma_{c}\pi$\\
         \hline
         $\Xi^{'}_{c}K$ & 0 & $-\sqrt{3}$ & $-\sqrt{2}$ \\
         $\Sigma_{c}\eta$ & $-\sqrt{3}$ & 0 & 0 \\
         $\Sigma_{c}\pi$ & $-\sqrt{2}$ & 0 & $-2$ \\
         \hline\hline
      \end{tabular}
   \end{table}

\begin{table}[!hbp]
      \caption{$(S=0,I=0)$}
         \begin{tabular}{c|cc}
         \hline\hline
           & $\Xi^{'}_{c}$K & $\Sigma_{c}\pi$\\
         \hline
         $\Xi^{'}_{c}$K & $-2$ & $-\sqrt{3}$\\
         $\Sigma_{c}\pi$ & $-\sqrt{3}$ & $-4$ \\
         \hline\hline
      \end{tabular}
  \end{table}

\begin{table}[!hbp]
      \caption{$(S=-1,I=3/2)$}
      \begin{tabular}{c|cc}
         \hline\hline
           & $\Xi^{'}_{c}\pi$ & $\Sigma_{c}\overline{K}$\\
         \hline
         $\Xi^{'}_{c}\pi$ & $1$ & $\sqrt{2}$\\
         $\Sigma_{c}\overline{K}$ & $\sqrt{2}$ & 0 \\
         \hline\hline
      \end{tabular}
 \end{table}

\begin{table}[!hbp]
      \caption{$(S=-1,I=1/2)$}
      \begin{tabular}{c|cccc}
         \hline\hline
           & $\Xi^{'}_{c}\pi$ & $\Xi^{'}_{c}\eta$ & $\Omega_{c}K$ & $\Sigma_{c}\overline{K}$\\
         \hline
         $\Xi^{'}_{c}\pi$ & $-2$ & 0 & $-\sqrt{3}$ & $\frac{1}{\sqrt{2}}$\\
         $\Xi^{'}_{c}\eta$ & 0 & 0 & $-\sqrt{3}$ & $\frac{3}{\sqrt{2}}$\\
         $\Omega_{c}K$ & $-\sqrt{3}$ & $-\sqrt{3}$ & $-2$ & 0\\
         $\Sigma_{c}\overline{K}$ & $\frac{1}{\sqrt{2}}$ & $\frac{3}{\sqrt{2}}$ & 0 & $-3$\\
         \hline\hline
      \end{tabular}
     \end{table}

\begin{table}[!hbp]
      \caption{$(S=-2,I=1)$}
      \begin{tabular}{c|cc}
         \hline\hline
           & $\Xi^{'}_{c}\overline{K}$ & $\Omega_{c}\pi$\\
         \hline
         $\Xi^{'}_{c}\overline{K}$ & $1$ & $\sqrt{2}$\\
         $\Omega_{c}\pi$ & $\sqrt{2}$ & 0 \\
         \hline\hline
      \end{tabular}
   \end{table}

\begin{table}[!hbp]
      \caption{$(S=-2,I=0)$}
      \begin{tabular}{c|cc}
         \hline\hline
           & $\Xi^{'}_{c}\overline{K}$ & $\Omega_{c}\eta$\\
         \hline
         $\Xi^{'}_{c}\overline{K}$ & $-1$ & $\sqrt{6}$\\
         $\Omega_{c}\eta$ & $\sqrt{6}$ & 0 \\
         \hline\hline
      \end{tabular}
\end{table}

\begin{table}[!hbp]
      \caption{$(S=-3,I=1/2)$\label{Table:coeff6l}}
         \begin{tabular}{c|c}
         \hline\hline
           & $\Omega_{c}\overline{K}$ \\
         \hline
         $\Omega_{c}\overline{K}$ & $2$\\
         \hline\hline
      \end{tabular}
\end{table}

\end{document}